\documentclass[10pt,journal,compsoc]{IEEEtran}

\usepackage{times,amsmath,epsfig}
\usepackage{multirow}
\usepackage{graphicx}
\usepackage{balance} 
\usepackage[ruled,vlined,linesnumbered]{algorithm2e}
\usepackage{comment}
\usepackage{subfigure}
\usepackage{color}
\usepackage[dvipsnames]{xcolor}
\usepackage{soul}
\usepackage[colorlinks=true,
	             linkcolor=black,
        	    	     urlcolor=black,
            	     citecolor=black]{hyperref}
\usepackage{lineno}
\usepackage{color,soul}
\usepackage{amsthm}
\usepackage{makecell}
\usepackage{enumerate}
\usepackage[normalem]{ulem}

\theoremstyle{plain}% default
\newtheorem{ex}{Example}

\definecolor{mycolor}{rgb}{0.77, 0.89, 0.99}

\newcommand*\kpp{\mathbf{\bar{|p|}}}
\newcommand*\ppb{\mathbf{\bar{|b|}}}

\ifCLASSOPTIONcompsoc
  \usepackage[nocompress]{cite}
\else
  \usepackage{cite}
\fi

\ifCLASSINFOpdf
\else
\fi
\begin{document}
%
% paper title
% Titles are generally capitalized except for words such as a, an, and, as,
% at, but, by, for, in, nor, of, on, or, the, to and up, which are usually
% not capitalized unless they are the first or last word of the title.
% Linebreaks \\ can be used within to get better formatting as desired.
% Do not put math or special symbols in the title.

\title{Schema-agnostic Progressive Entity Resolution}

\author{Giovanni Simonini, George Papadakis, Themis Palpanas, and Sonia Bergamaschi% <-this % stops a space
\IEEEcompsocitemizethanks{\IEEEcompsocthanksitem G. Simonini and S. Bergamaschi are with the Department of Engineering ``Enzo Ferrari'' of the University of Modena and Reggio Emilia, Italy.
\newline
E-mail: \{simonini, sonia\}@unimore.it
\IEEEcompsocthanksitem George Papadakis is with the University of Athens,
Greece.
\newline
E-mail: gpapadis@di.uoa.gr 
\IEEEcompsocthanksitem Themis Palpanas is with the Paris Descartes University, France.
\newline
E-mail: themis@mi.parisdescartes.fr
}% <-this % stops an unwanted space
\thanks{Manuscript received date; revised date.}}

%\markboth{}%
%{}

\IEEEtitleabstractindextext{%
\begin{abstract}
%\color{NavyBlue}
Entity Resolution (ER) is the task of finding entity profiles that correspond to the same real-world entity.
Progressive ER aims to efficiently resolve large datasets when limited time and/or computational resources are available.
In practice, its goal is to provide the best possible partial solution by approximating the optimal comparison order of the entity profiles.
%When limited time or computational resources are available, Progressive ER aims to provide the best possible partial solution by approximating the optimal comparison order of the entity profiles.
So far, Progressive ER has only been examined in the context of structured (relational) data sources, as the existing methods rely on schema knowledge
to save unnecessary comparisons: they restrict their search space to similar entities with the help of schema-based blocking keys (i.e., signatures that represent the entity profiles).
As a result, these solutions are not applicable in Big Data integration applications, which involve large and heterogeneous datasets, such as relational and RDF databases, JSON files, Web corpus etc.
%To cover this gap, we propose new schema-agnostic Progressive ER methods, which do not require schema information for their blocking keys, thus applying to heterogeneous data sources of any schema variety.
To cover this gap, we propose a family of schema-agnostic Progressive ER methods, which do not require schema information, thus applying to heterogeneous data sources of any schema variety.
First, we introduce two na\"ive schema-agnostic methods, showing that 
straightforward solutions exhibit a poor performance that does not scale well
to large volumes of data.
Then, we propose four different advanced methods.
%The first two are based on a sorted list of entity profiles, exploiting the \textit{similarity} of the blocking keys, while the third is based on a graph of entity profiles, exploiting the \textit{equality} of the blocking keys.
%The former achieve the highest performance for structured data, and the latter for semi-structured ones.
Through an extensive experimental evaluation over 7 real-world, established datasets, we show that all the advanced methods outperform to a significant extent both the na\"ive and the state-of-the-art schema-based ones.
We also investigate the relative performance of the advanced methods, providing guidelines on the method selection.
\end{abstract}
\vspace{-7pt}
% Note that keywords are not normally used for peerreview papers.
}

\maketitle

\IEEEdisplaynontitleabstractindextext

\IEEEpeerreviewmaketitle

\section{Introduction}
\label{sec:introduction}

When dealing with heterogeneous data, real-world entities may have different
representations; for instance, they can be records in a relational database,
sets of RDF triples,  JSON objects, text snippets in a web corpus, etc.
We call \textit{entity profile} (or simply \textit{profile}) each representation
of a real-world \emph{entity} in data sources. The task of identifying
different profiles that refer to the same real-world entity is called 
Entity Resolution (\textsf{ER}) and constitutes a critical process that has many
applications in areas such as Data Integration, Social Networks, and Linked Data
\cite{DBLP:series/synthesis/2015Christophides,DBLP:series/synthesis/2015Dong, DBLP:journals/pvldb/GetoorM12}.
%\note{motivation}

\textsf{ER} can be distinguished into two broad
categories~\cite{DBLP:journals/tkde/PapenbrockHN15,DBLP:journals/tkde/WhangMG13}:
\emph{(i)} \textsf{Off-line} or \textsf{Batch ER}, which aims to provide a
\textit{complete solution}, after all processing is terminated, and \emph{(ii)}
\textsf{On-line} or \textsf{Progressive ER}, which aims to provide the best
possible \textit{partial solution}, when the response time, or the available
computational resources are limited.
The latter is driven by modern \textit{pay-as-you-go} applications that do not
require the complete solution to produce useful results.
%~\cite{DBLP:journals/tkde/WhangMG13}.
%\note{Giovanni: for instance,...}
%{\color{red}George: should we refer to an Progressive ER application as a
%motivation, here?}

\textsf{Progressive ER} is becoming increasingly
important~\cite{DBLP:conf/pods/GolshanHMT17,DBLP:journals/tkde/PapenbrockHN15,DBLP:journals/tkde/WhangMG13},
as the number of data sources and the amount of
available data multiply. For example, the number of high-quality
HTML tables on the Web is in the hundreds of millions, while the
Google dataset search system alone has indexed $\sim$26 billion datasets
\cite{DBLP:conf/pods/GolshanHMT17}. This huge volume of data can only be
resolved in a pay-as-you-go fashion, especially for applications with strict
time requirements, e.g., the catalog update in large online
retailers that is carried out every few hours\footnote{www.nchannel.com/blog/challenges-ecommerce-catalog-management/}.
Most importantly, Web data abound in highly diverse, multilingual, noisy
and incomplete schemata to such an extent that it is practically
impossible to unify them under a global schema
\cite{DBLP:conf/pods/GolshanHMT17}. Inevitably, this unprecedented variety
renders schema-based progressive methods inapplicable to Web data. For these
reasons, 
% while at the same time the
%relevant applications have strict time requirements.
%(e.g., shopping catalogs)  (e.g., on the Web)
%(e.g., updating catalogs every few hours for large online retailers
%In this paper, 
we propose novel, schema-agnostic \textsf{Progressive ER}
methods that go beyond the current state-of-the-art approaches in all respects -
we outperform them significantly even when reliable schema
information is available.
% {\bf ??? motivation para: if you are not convinced, it is probably not good
% enough... ???}

%\begin{figure}[t!]\centering
%	\includegraphics[width=0.49\textwidth]{images/profiles}
%	\vspace{-10pt}
%	\caption{A portion of a \textit{datalake} where an imaginary company stores all the information about its clients as structured, semi-structured, and unstructured data.}
%	\vspace{-15pt}
%	\label{fig:profiles}
%\end{figure}

%\vspace{2pt}
\noindent
\textbf{Progressive Methods.}
A core characteristic of the existing methods for \textsf{Progressive ER}
% , so far proposed, 
is that they rely on \textit{blocking} in order to scale to
large datasets
\cite{DBLP:journals/tkde/PapenbrockHN15,DBLP:journals/tkde/WhangMG13}. 
Blocking is a typical pre-processing step for \textsf{Batch ER} that aims to
index together \textit{likely-to-match} profiles into buckets (called
\textit{blocks}), according to an indexing criterion (called \textit{blocking key}).
Thus, comparisons are limited to pairs of profiles that co-occur in at
least one block, avoiding the quadratic complexity of the na\"ive
\textsf{ER} solution, which compares every profile with all others. In this
way, progressive methods generate on-line the most promising \textit{pairs
of profiles} to be compared {by a \textit{match function},
i.e., a (usually) binary function that decides whether two given profiles are matching, or not.}

In fact, progressive methods 
use blocking to generate on-line pairs of profiles in decreasing order of
matching likelihood. So far, however, they have been exclusively combined with
\textit{schema-based} blocking \cite{DBLP:journals/tkde/PapenbrockHN15,DBLP:journals/tkde/WhangMG13}, 
which is specifically crafted for structured (relational) data. 
That is, they rely on schema knowledge in order to build blocks of low noise
and high discriminativeness, 
%, achieving high recall and precision; in case the resulting recall is inadequate, multiple schema-based blocking keys are used.
%The implicit assumption is that all input records abide by a schema with
% attributes of known quality.
assuming implicitly that all input records abide by a schema with
attributes of known quality.

%\vspace{2pt}
\noindent
\textbf{Limitations of Existing Approaches.} The existing progressive methods
suffer from the following major drawbacks:

%\vspace{2pt}
\noindent
\emph{(1)} In practice, their fundamental assumption that schema %(quality)
is a-priori known holds for a small portion of the data we would like to
handle. For instance, Web data typically comprises large, semi-structured,
heterogeneous entities that manifest two main challenges of Big Data
\cite{DBLP:series/synthesis/2015Christophides,DBLP:series/synthesis/2015Dong}:
\emph{(i)} \textit{Volume}, as they involve millions of entities that are
described by billions of statements, and \emph{(ii)} \textit{Variety}, since
their descriptions entail thousands of different attribute names.
More generally, in a Big Data integration scenario, schema-alignment is too
expensive and time consuming when multiple heterogeneous data sources are
involved \cite{DBLP:conf/pods/GolshanHMT17,DBLP:series/synthesis/2015Dong}, thus
yielding a prohibitively high cost for pay-as-you-go applications.

\begin{figure}[t!]\centering
	\includegraphics[width=0.47\textwidth]{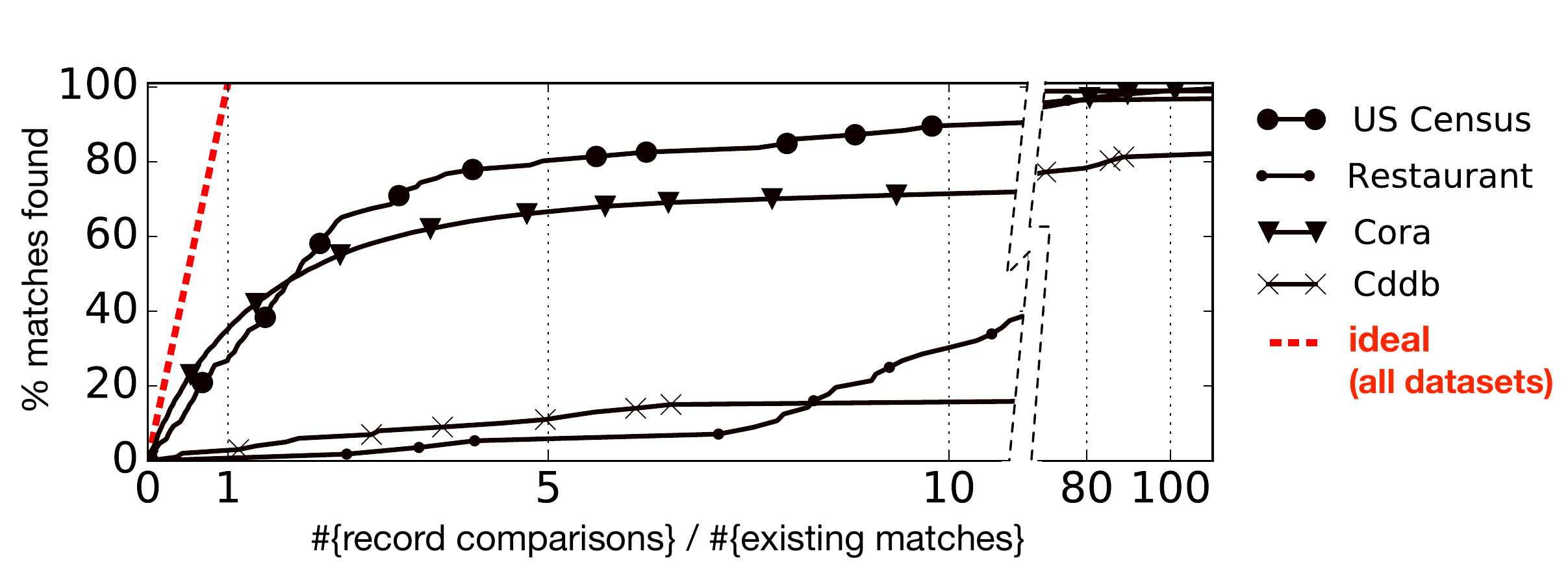}
	\vspace{-8pt}
	\caption{The performance of Progressive Sorted Neighborhood on 4 real-world structured datasets.
	}
	\vspace{-12pt}
	\label{fig:ex01}
\end{figure}

%\vspace{2pt}
\noindent
\emph{(2)} Even when the schema assumption holds, there is plenty of room for
improving the performance of existing schema-based progressive methods. 
We demonstrate this
in Figure \ref{fig:ex01} over four established,
real-world and diverse datasets:
the state-of-the-art schema-based method, Progressive Sorted Neighborhood
(\textsf{PSN})
\cite{DBLP:journals/tkde/PapenbrockHN15,DBLP:journals/tkde/WhangMG13}, finds 
only $\sim$60\% and $\sim$85\% of all matches for \texttt{Cora} and \texttt{US
Census}, respectively, after executing 10 times the number of comparisons
required by the optimal algorithm to identify 100\% of the matches (i.e., 1
comparison per pair of duplicates).
For the rest of the datasets, the performance is also far from optimal:
for \texttt{Restaurant}, \textsf{PSN} identifies 
almost all matches only after performing 2 orders of magnitude more
comparisons than the optimal algorithm, while for \texttt{Cddb}, it detects less than
80\% of all matches with the same (excessive) number of comparisons.

%\vspace{2pt}
\noindent
\textbf{Our Contributions.}
We propose novel and unsupervised methods for \textsf{Progressive ER} that
inherently address the Variety of Big Data: they operate in a schema-agnostic
fashion, which overrides the
% there is no 
need to search for and identify highly discriminative attributes,
% and, thus,
%and consequently, there is no 
% the need for 
rendering schema knowledge unnecessary.
%  either. In this way,
% they inherently address the Variety of Big Data.
Our methods are also more
effective in addressing the Volume of Big Data, since they 
identify matches earlier than the top-performing schema-based method. They
actually go beyond the state-of-the-art in \textsf{Progressive ER}
by introducing and exploiting \textit{redundancy}, i.e., by associating every
profile with multiple blocking keys. Instead, existing schema-based progressive
methods typically rely on highly discriminative attributes, which yield
\textit{redundancy-free keys} such that two profiles appear together
in at most one block.
% In contrast, our methods associate every profile with multiple blocks.

More specifically, our redundancy-based methods rely on two principles.
The first one is called \textit{\textbf{similarity principle}}, as it
%  these
% methods 
assumes that any two matching profiles have blocking keys that are closer in 
alphabetical order than those of non-matching ones.
The second one is called \textit{\textbf{equality principle}}, since it
%  these methods 
assumes that the matching likelihood of any two profiles is proportional to the
number of blocks they share.
Both principles have been successfully applied in \textsf{Batch
ER}~\cite{DBLP:journals/pvldb/0001APK15}, but their application to the progressive
context is non-trivial, as we show empirically. For this
reason, we introduce more advanced methods for every principle. 

Through an exhaustive experimental evaluation over 7 well-known datasets, we
verify that similarity-based methods excel in structured datasets, outperforming
even the state-of-the-art schema-based progressive method. These datasets
typically involve a large portion of textual information, which provides 
reliable matching evidence when sorted alphabetically. In contrast,
our equality-based method is the top-performer over semi-structured datasets
(e.g., RDF data); it can exploit the semantics of the URIs that abound in this
type of datasets, disregarding the useless information of URI prefixes, which
introduce high levels of noise when sorted alphabetically.

% , the same approach
% cannot be applied 
% The advantage of our methods is that there is no need to search for and identify
% highly discriminative attributes, and consequently, there is no need for schema
% knowledge either.
% As we demonstrate in our experimental evaluation, this approach leads to better
% performance, as well.

On the whole, we make the following contributions:
%\begin{itemize}
%[leftmargin=*]

\noindent
$\bullet$ 
%\item 
We introduce a \textit{schema-agnostic} approach to \textsf{Progressive
ER}, which inherently addresses the Variety of Big Data.

\noindent
$\bullet$
%\item 
We demonstrate that adapting existing schema-based methods to
schema-agnostic \textsf{Progressive ER}  is a non-trivial task: we introduce 2 na\"ive,
schema-agnostic methods,
% in Section \ref{sec:naiveSolutions} and Appendix B,
showing experimentally that they fail to address the Volume issue of Big Data.

\noindent
$\bullet$ 
%\item 
We present 4 novel advanced, schema-agnostic progressive  methods, 
%(Section \ref{sec:approach}), 
which 
%successfully 
address both the Volume and the
Variety of Big Data.
They are classified in two categories: those based on a sorted list of profiles,
leveraging the \emph{similarity principle},
% (Section \ref{sec:sn-methods}), 
and those based on a graph of profiles, leveraging the \emph{equality
principle}.
%(Section \ref{sec:bg-methods}).
% They are classified in two categories: those based on the \emph{similarity} of
% blocking keys and the resulting sorted list of profiles (Section
% \ref{sec:sn-methods}) and those based on the the \emph{equality} of blocking
% keys and the graph extracted from the corresponding set of blocks (Section
% \ref{sec:bg-methods}).

\noindent
$\bullet$
%\item 
We perform a series of experiments over 7 established, real-world
datasets, 
%(Section \ref{sec:experiments}), 
verifying experimentally the superiority
of our methods over the existing schema-based state-of-the-art
method, both in terms of effectiveness and time efficiency.
We also investigate the relative performance of our methods, highlighting the
top-performing ones, and providing guidelines for method selection.

%\end{itemize}

The rest of the paper is structured as follows: Section \ref{sec:relatedWork}
discusses the main works in the literature, while Section
\ref{sec:preliminaries} describes the background of our methods. We present two
na\"ive schema-agnostic solutions to \textsf{Progressive ER} in Section
\ref{sec:naiveSolutions}, and four advanced ones in Section \ref{sec:approach}.
We elaborate on our extensive experimental evaluation in Section
\ref{sec:experiments} and conclude the paper in Section \ref{sec:conclusion},
along with directions for future work.

% We conclude the paper in Section \ref{sec:conclusion}, while in the Appendix, we summarize the notation used throughout the paper and present a taxonomy of all
% discussed methods to facilitate their understanding.

\vspace{-3pt}
\section{Related Work}
\label{sec:relatedWork}
%{\color{blue}ER is a well known problem for data integration and data cleaning (see \cite{DBLP:journals/pvldb/GetoorM12} for a survey)}

\noindent \textbf{Schema-based Progressive Methods}.
The state-of-the-art progressive method is \textit{Progressive Sorted
Neighborhood} (\textbf{\textsf{PSN}}) \cite{DBLP:journals/tkde/PapenbrockHN15,
DBLP:journals/tkde/WhangMG13}.
Based on Batch Sorted Neighborhood \cite{DBLP:conf/sigmod/HernandezS95}, it
associates every profile with a schema-based blocking key. Then, it produces a
\textit{sorted list of profiles} by ordering all blocking keys alphabetically.
Comparisons are progressively defined through a sliding
window, $w$, whose size is \textit{iteratively incremented}: 
initially, all profiles in consecutive positions ($w$=1) are compared, starting
from the top of the list; then, all profiles at distance $w$=2 are compared and
so on and so forth, until the processing is terminated.

However, the performance of \textsf{PSN} depends heavily on the attribute(s)
providing the schema-based blocking keys that form the sorted
list(s) of profiles. In case of low recall, the entire process is repeated,
using multiple blocking keys per profile. As a result, \textsf{PSN} requires
domain experts, or supervised learning on top of labeled data in order to
achieve high performance. 
In contrast, our methods are completely unsupervised and schema-agnostic.

Two more schema-based methods were proposed in
\cite{DBLP:journals/tkde/WhangMG13}:
\textit{Hierarchy of Record Partitions} (\textbf{\textsf{HRP}}) and
\textit{Ordered List of Records} (\textbf{\textsf{OLR}}).
The main idea of \textsf{HRP} is to build a hierarchy of blocks, such that the
matching likelihood of two profiles is proportional to the level in which they
appear together for the first time: the blocks at the bottom of
the hierarchy contain the profiles with the highest matching likelihood, and vice
versa for the top hierarchy levels. Thus, the hierarchy of blocks can be
progressively resolved, level by level, from the leaves to the root. This
approach has been improved in the literature in two ways: \emph{(i)}
\textsf{OLR} exploits this hierarchy in order to produce a list of records
sorted by their likelihood to produce matches, 
% trading . Compared to
% \textsf{HRP}, \textsf{OLR} trades 
involving a lower memory consumption than \textsf{HRP} at the cost of a
slightly worse performance. \emph{(ii)} A schema-based variation of \textsf{HRP}
is adapted to the MapReduce parallelization framework for even higher efficiency
in \cite{DBLP:conf/icde/AltowimM17}. It divides every block into a
hierarchy of child blocks and uses an advanced strategy for optimizing their
parallel processing.

However, both \textsf{HRP} and \textsf{OLR} are difficult
to apply in practice. The hierarchies that lie at their core can be generated only
when the distance of two records can be naturally estimated through a certain
attribute (e.g., product price) \cite{DBLP:journals/tkde/WhangMG13}.
The number of the hierarchy layers, $L$, has to be determined
a-priori, along with $L$ similarity thresholds and the similarity measure
that compares attribute values. 
Moreover, they both exhibit a performance inferior to \textsf{PSN}~\cite{DBLP:journals/tkde/WhangMG13}. 
For these reasons, we do not consider these two methods any further.

%\color{blue}
Altowim et al. \cite{DBLP:journals/pvldb/AltowimKM14} propose a
progressive \textit{joint} solution in the context of multiple
datasets containing different entity types.
% for \textit{joint} \textsf{ER} \cite{DBLP:conf/icde/WhangG12}, that is to
Similarly, in joint \textsf{ER} \cite{DBLP:conf/icde/WhangG12}, the result on
one dataset can be exploited to resolve the others.
As an example, let us consider a joint \textsf{ER} on a \textit{movie dataset}
and on an \textit{actor dataset}: discovering matches among actors can help to
determine whether two movies associated to those actors are matching too (and
vice versa).
% if two apparently different movies are associated to the same list of actors,
% then it is likely that the two movies are actually the same; thus,
Both approaches, though, are only applicable to Relational
\textsf{ER}.

\begin{figure}[t!]
\centering
	\includegraphics[width=0.49\textwidth]{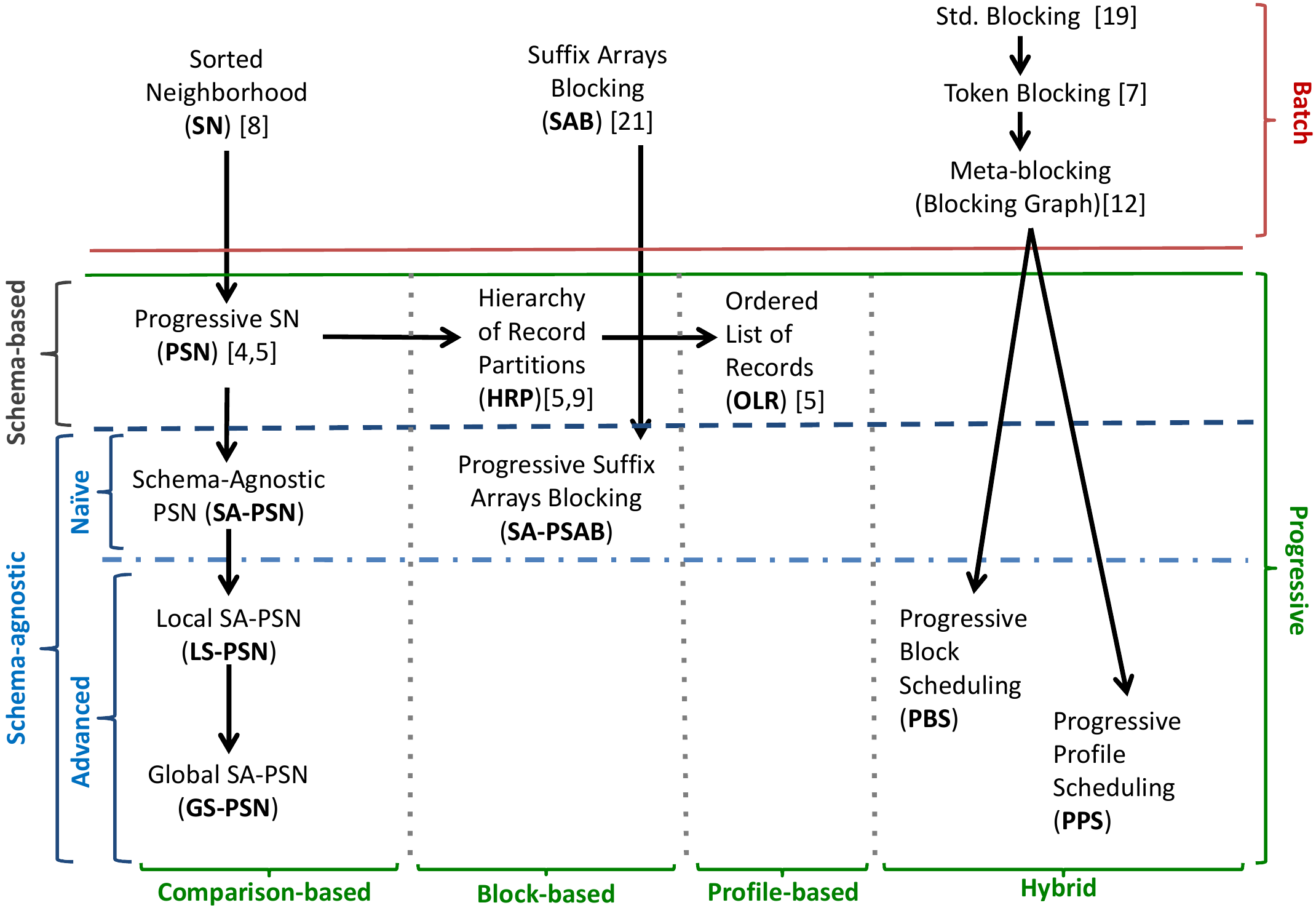}
	\vspace{-20pt}
	\caption{
	Overview of all discussed progressive methods.
	$A {\rightarrow} B$ means that method $B$ builds on method $A$ in order to offer a new functionality.
	}
	\label{fig:taxonomy}
	\vspace{-12pt}
\end{figure}

\vspace{2pt}
\noindent
\textbf{Taxonomy of Progressive Methods.}
We now present a taxonomy that organizes the existing progressive methods and
those presented in the following with respect to the granularity of their
functionality. This results in four categories, which generalize
the hint types discussed in \cite{DBLP:journals/tkde/WhangMG13}:
%Comparison-based, Block-based, and Entity-based.

%\noindent
%\begin{enumerate}
\emph{(1)}
%\item 
\textit{Comparison-based methods} provide a list of profile pairs
(i.e., comparisons) that are sorted in descending order of matching likelihood
(from the highest to the lowest one). With every method call, these profile
pairs are then emitted, one at a time, following that ordered list.
This category is a generalization of the category \textit{``sorted
list of record pairs''} \cite{DBLP:journals/tkde/WhangMG13} and includes the methods \textsf{PSN
\cite{DBLP:journals/tkde/PapenbrockHN15, DBLP:journals/tkde/WhangMG13}},
\textsf{SA-PSN} (see Section \ref{sec:sapsn}), \textsf{LS-PSN} (see Section
\ref{sec:lspsn}), and \textsf{GS-PSN} (see Section \ref{sec:gspsn}).

%\noindent
\emph{(2)}
%\item 
\textit{Block-based methods} produce a list of blocks that are sorted in
descending order of the likelihood that they include duplicates among their
profiles.
In every call, all the comparisons for each block are generated,
one block at a time, following that ordered list;~all comparisons in the same
block have the same matching likelihood. This is a generalization of the
category \textit{``hierarchy of record partitions''}
\cite{DBLP:journals/tkde/WhangMG13} and includes the homonymous method
\textsf{HRP} \cite{DBLP:journals/tkde/WhangMG13} together with \textsf{SA-PSAB}
(see Section \ref{sec:sapsab}).

\emph{(3)}
%\item
\textit{Profile-based methods} provide a list of profiles that are sorted
in descending order of duplication likelihood. Then, in every call,
all comparisons of every entity are generated, one entity at a time,
following that ordered list. This category is a generalization of the category
\textit{``ordered list of records''} \cite{DBLP:journals/tkde/WhangMG13} and
includes the homonymous method \textsf{OLR} \cite{DBLP:journals/tkde/WhangMG13}.

\emph{(4)}
\textit{Hybrid methods} combine characteristics from two or all of the
previous categories. This category includes \textsf{PBS}
(see Section \ref{sec:pbsMethod}), which involves both block- and comparison-based
characteristics, as well as \textsf{PPS} (see Section
\ref{sec:pps}), which combines comparison-based
characteristics with profile-based ones.

%\end{enumerate}

We illustrate our taxonomy in Figure \ref{fig:taxonomy}, where every
column corresponds to a different type of granularity (horizontal axis). 
% Note that we have also
% added a fourth column, which corresponds to the category of \textit{hybrid
% methods}. These are methods that combine characteristics from two or all of the
% previous categories. For example, it includes \textsf{PBS} (see Section
% \ref{sec:pbsMethod}), which involves both block- and comparison-based
% characteristics, as well as \textsf{PPS} (see Section
% \ref{sec:pps}), which combines comparison-based
% characteristics with profile-based ones.
% 
% Note that Figure \ref{fig:taxonomy} maps all progressive methods discussed
% in this work to the corresponding granularity category (horizontal axis). 
On
the vertical axis, we consider the relation of every progressive method to
schema knowledge, with the topmost part corresponding to batch methods.
% columns classify the methods on the basis of their granularities.
Every arrow from method $A$ to method $B$ means that $B$ extends $A$ to
offer a new functionality. For instance, \textsf{PBS} and \textsf{PPS}
are based on the Blocking Graph, which is the core data structure of
Batch Meta-blocking \cite{DBLP:conf/edbt/0001PPK16}.

%\color{blue}
\vspace{2pt}
\noindent
\textbf{Crowdsourced (or Oracle) Methods}.
In \textit{Crowdsourced} \textsf{ER} \cite{DBLP:journals/pvldb/WangKFF12}, humans are
asked to label candidate profile pairs as either matching or non-matching, i.e.,
they are asked to behave like a binary \textit{match function}.
% In such a context, it is reasonable to assume this crowdsourced match function
% to be \textit{transitive}.
Such a function is typically assumed to be \textit{perfect}
(i.e., being equivalent to an \textit{oracle}
\cite{DBLP:journals/pvldb/VesdapuntBD14}) and \textit{transitive}
\cite{DBLP:conf/sigmod/WangLKFF13}.
For example, given three profiles ($p_1$, $p_2$, $p_3$), if the crowd finds
that $p_1$ matches with $p_2$, and $p_2$ with $p_3$, then the
comparison between $p_1$ and $p_3$ is not crowdsourced, but is
automatically deduced as a match.
%the crowd is surely find $p_1$ matching with $p_3$.
Progressive crowdsourced methods \cite{DBLP:conf/sigmod/WangLKFF13,
DBLP:journals/pvldb/VesdapuntBD14, DBLP:journals/pvldb/FirmaniSS16} exploit this
transitivity to maximize the progressive recall of \textsf{ER}.
In this work, though, we propose general methods for \textsf{Progressive ER}
that are independent of the employed match function, i.e., we do not assume the
match function to be transitive, nor to be perfect|a setting that is common for
(non-crowdsourced) match functions \cite{DBLP:journals/vldb/BenjellounGMSWW09}.
We exclusively consider progressive methods
that define a static processing order, without relying on the feedback of
the ``match function'' to dynamically re-adjust it, as in crowdsourced methods
or the ``look-ahead strategy'' that lies at the core of
\cite{DBLP:journals/tkde/PapenbrockHN15}.

% Yet, our methods could be combined with the current
% state-of-the-art crowdsourced one, which is presented in
% \cite{DBLP:journals/pvldb/FirmaniSS16}.
% %  In fact, Firmani et al.
% % \cite{DBLP:journals/pvldb/FirmaniSS16} 
% Before submitting the first record pair to the crowd, this approach 
% % assumes the existence of 
% builds a list of record pairs sorted in
% decreasing \textit{matching likelihood}.
% %  that is built before submitting the
% % first record pair to the crowd.
% To assess this matching likelihood,
% % of the record pairs, 
% % following
% % \cite{DBLP:journals/pvldb/WangKFF12}, Firmani et al.
% % \cite{DBLP:journals/pvldb/FirmaniSS16} 
% it computes the string similarity (e.g.,
% Jaro-Winkler or Jaccard) of all possible record pairs|a very similar
% pre-processing step is required by
% \cite{DBLP:journals/pvldb/WangKFF12}, \cite{DBLP:journals/pvldb/VesdapuntBD14}, \cite{DBLP:conf/sigmod/WangLKFF13} and \cite{DBLP:journals/tkde/SongLH17}.
% %\footnote{\color{red}This method is not crowdsource-based, but relies on this pre-processing step as well.}.
% This is a prohibitively expensive pre-processing step when a low latency response is required, even if a blocking method is used to avoid the all-pairs
% comparisons.
% As an alternative, our methods could be employed to generate the sorted list of
% record pairs efficiently (instead of using them directly with match functions).
%We do not investigate this direction in our work.
\vspace{-3pt}
\section{Preliminaries}
\label{sec:preliminaries}

\begin{figure*}[th!]	
	\centering
	\includegraphics[width=1.0\textwidth]{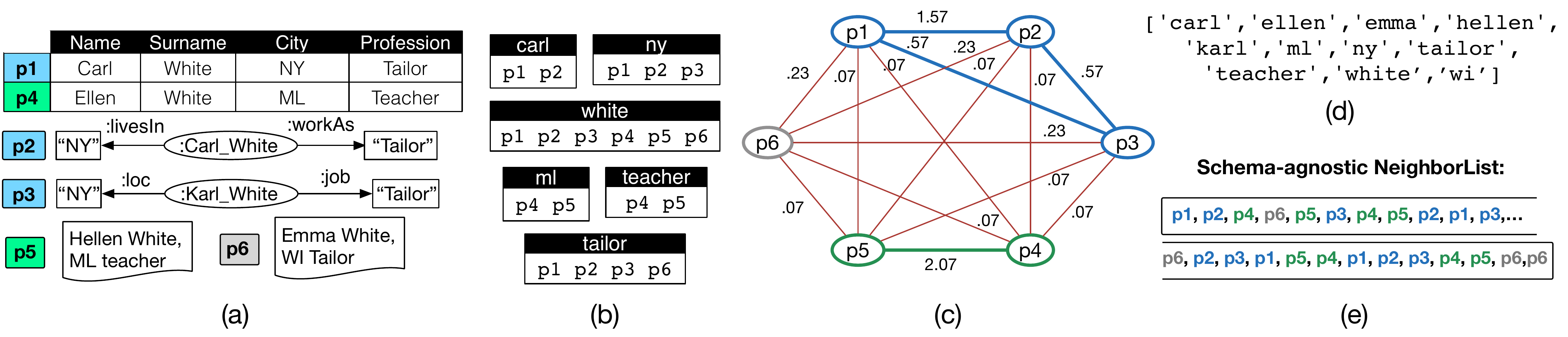}
	\vspace{-20pt}
	\caption{(a) A set of profiles $P$ that is extracted from a \textit{data lake}
	with a variety of data formats: structured/relational data ($p_1$, $p_4$),
	semi-structured/RDF data ($p_2$, $p_3$) and unstructured/free-text data ($p_5$,
	$p_6$).
	Note that $p1{\equiv}p2{\equiv}p3$ and $p4{\equiv}p5$.
	(b) The block collection $B$ derived from $P$ by applying Token Blocking
	\cite{DBLP:conf/wsdm/PapadakisINF11} to its profiles.
	(c) The Blocking Graph derived from $B$, with every edge representing a profile
	comparison that is weighted by the \textsf{ARCS} function.
	%and the average weight of the incident edges of each node is listed in the box.
	(d) The sorted list of attribute value tokens that appear in the profiles of
	$P$. (e) The corresponding schema-agnostic Neighbor List.
	}
	\vspace{-15pt}
	\label{fig:example}
\end{figure*}

At the core of \textsf{ER} lies the notion of \textit{entity
profile} (or simply \textit{profile}), which constitutes a uniquely identified
set of attribute name-value pairs.
An individual profile is denoted by $p_i$, with $i$ standing for its id in a
\textit{profile collection} $P$.
Two profiles $p_i, p_j \in P$ are called \textit{duplicates} or \textit{matches}
($\mathbf{p_i\equiv p_j}$) if they represent the same real-world entity.

Depending on the input data, \textsf{ER} takes two forms
\cite{DBLP:series/synthesis/2015Christophides,DBLP:series/synthesis/2015Dong}:
\emph{(1)} \textsf{Clean-clean ER} receives as input two duplicate-free, but
overlapping profile collections, $P_1$ and $P_2$,
 and returns as output all pairs of duplicate
profiles they contain, $P_1 \cap P_2$.
\emph{(2)} \textsf{Dirty ER} takes as input a single profile collection that
contains duplicates in itself and produces a set of equivalence clusters, with
each one corresponding to a distinct profile.

To scale \textsf{ER} to large data collections, \textit{blocking} is employed to cluster
similar profiles into \textit{blocks} so that it suffices to consider
comparisons among the profiles of every block \cite{DBLP:journals/tkde/Christen12}.
Each profile is \textit{indexed} into blocks according to one or more criteria
called \textit{blocking keys}.
If a blocking key depends on the schema(ta) of the data,
% source(s), 
we call it \textit{schema-based}, otherwise \textit{schema-agnostic}.

An individual block is symbolized by $b_i$, with $i$ corresponding to its id.
The size of $b_i$ (i.e., the number of profiles it contains) is denoted by
$|b_i|$ and its cardinality (i.e., the number of comparisons that it yields) by
$\|b_i\|$.
{For instance, in Figure~\ref{fig:example}(b), $|b_{tailor}|{=}4$ 
%(as the number of profiles) 
and 
$\|b_{tailor}\| {=} {4 \choose 2} {=} 6$.
 %(as, the number of comparisons entailed)
}
A set of blocks $B$ is called \textit{block collection}, with $|B|$ standing for
its size (i.e., total number of blocks) and $\|B\|$ for its aggregate
cardinality (i.e., the total number of comparisons entailed by $B$):
% \begin{equation*}
$\|B\|$=$\sum_{b_i\in B}\|b_i\|$.
%\end{equation*}
The set of blocks associated with a specific profile $p_i$ is denoted by $B_i$,
and the \textit{average number of profiles per block} by
% $\mathbf{\bar{|p|}}$ and $\bar{|b|}$, respectively:
% \begin{equation*}
% $\kpp=\sum_{b \in B} {|b|}/{|P|}$, and
% \hspace{15pt}
$\ppb$=$\sum_{b\in B} {|b|}/{|B|}$.
% \end{equation*}
The comparison between profiles $p_i$ and $p_j$ is symbolized by
$c_{ij}$.
% while $\kpp$ stands for the \textit{average number of name-value pairs
%per profile}.

\vspace{-10pt}
\subsection{Progressive ER}
\label{sec:per_methods}

In \textsf{Batch ER}, the profile comparisons entailed in block collection $B$
are executed without a specific order. Let $T_o$ be the overall time required
for performing \textsf{Batch ER} on $B$. Based on $T_o$,
%$B$ is completely defined before starting the comparisons.
% In the following, we formally define \textsf{Progressive ER} starting the notion of \textsf{Batch ER}.
%
%Formally, a 
\textsf{Progressive ER}
% method 
is formally defined by two requirements
\cite{DBLP:journals/tkde/PapenbrockHN15,DBLP:journals/tkde/WhangMG13}:
%\begin{itemize}[leftmargin=0pt]
  %[leftmargin=*]
  %\emph{(1)} 
 %\item 
 %\noindent

$\bullet$ 
 \textit{Improved Early Quality.} 
  If both \textsf{Progressive} and \textsf{Batch ER} are applied to 
  $B$ and terminated at the same time $t {\ll} T_o$, then the former
  should detect significantly more matches than the latter.
  %The performance at time $t \ll T_o$ should be as close as possible to the performance of the at time $T_o$, where $T_o$ is the overall time that is required for performing \textsf{Batch ER} on the given entity profiles.

  %\emph{(2)} 
 %\item 
 %\noindent
$\bullet$ 
 \textit{Same Eventual Quality.} The result produced at time $T_o$
  by \textsf{Progressive} and \textsf{Batch ER} should be identical. Even though
  progressive methods rarely run for so a long time as $T_o$, this
  requirement ensures their correctness, verifying that they yield the
  exact same outcome as batch methods.
%\end{itemize}
%%%%%%%%%%%%%%%%%%%%%%%%%%%%%%%%%%%%%%%%
%  old related work
%%%%%%%%%%%%%%%%%%%%%%%%%%%%%%%%%%%%%%%%

%We now present the main schema-based techniques for \textsf{Progressive ER} that cover all three categories of functionality granularity.

%We examine every category separately, following the same order of presentation as in \cite{DBLP:journals/tkde/WhangMG13}, since the methods of the later types build on those of the earlier ones.

%\subsubsection{Methods functionality}
In the following, we break the functionality of progressive
%relevant 
methods into two phases:
%\begin{enumerate}[i)]
%[leftmargin=*]

\emph{(1)}
%\item 
The \textbf{initialization phase} takes as input the
profiles to be resolved, builds the data structures needed for their processing,
and processes them to produce the overall best comparison.

\emph{(2)}
%\item 
The \textbf{emission phase} returns the next best
comparison from a list of candidates that are ranked in
non-increasing order of matching likelihood. In other words, 
%the emission phase 
it identifies the remaining pair of profiles that has the highest
matching likelihood.
% of matching.
%\end{enumerate}
%\noindent 

By definition, the initialization phase is activated just once, while the
emission phase is repeated whenever a new comparison is requested for
processing.

\vspace{-3pt}
\subsection{Core Data Structures}
We now describe two fundamental data structures for our progressive methods:
the \textit{Blocking Graph} and the \textit{Neighbor List}.
Every method discussed in the following has at its core either the former or the
latter. Note that both data structures are known from the
literature, sometimes with different names (e.g., the Neighbor List is called
\textit{sorted list of records} in~\cite{DBLP:journals/tkde/WhangMG13}).

%For each data structure, we highlight the main shortcomings of the existing methods employing it.

%%%%%%%%%%%%%%%%%%%%%%%%%%%%%%%%%%%%%%%%%%%%%%%%%%%%
%
%
%
% 			BLOCKING GRAPH
%
%
%
%%%%%%%%%%%%%%%%%%%%%%%%%%%%%%%%%%%%%%%%%%%%%%%%%%%%%

%\begin{figure}[t!]
%	\centering
%	\begin{tabular}{l}
%		\includegraphics[width=0.4\textwidth]{images/bg}\\
%	\hspace{0.35in}	(a)              \hspace{1.2in}                     (b)
%	\end{tabular}
%	%\vspace{-24pt}
%	\caption{A block collection (a) derived from the profiles in Figure \ref{fig:profiles} applying Token Blocking, and the corresponding Blocking Graph (b): the weights on the edges are derived with \textsf{ARCS}; while the weight in the box, for each node, represents the average weight of its incident edges.
%	Red edges correspond to non-matching comparisons.}
%	\vspace{-10pt}
%	\label{fig:bg}
%\end{figure} 

\vspace{2pt}
\noindent\textbf{Blocking Graph} |
This data structure lies at the core of \textsf{Batch Meta-blocking}
\cite{DBLP:series/synthesis/2015Christophides,DBLP:series/synthesis/2015Dong,DBLP:conf/edbt/0001PPK16,
DBLP:journals/pvldb/SimoniniBJ16}, which aims at restructuring an existing block
collection $B$ into a new one $B'$ that has similar recall, but significantly
higher precision than $B$. Meta-blocking relies on the assumption that the
matching likelihood of any two profiles is analogous to their degree of
co-occurrence in a block collection.
This means that $B$ has to be generated by a blocking method that yields
\textbf{redundancy-positive blocks}, where the similarity of two profiles 
is proportional to the number of blocks they share.  
%associating every profile with multiple blocks so that similar pairs of profiles share multiple blocks.

Based on redundancy, which is common for schema-agnostic blocking
methods \cite{DBLP:conf/edbt/0001PPK16}, 
Meta-blocking represents the block collection as a \textit{blocking graph}. 
This is an undirected weighted graph $\mathcal{G}_B(V_B, E_B)$,
where $V_B$ is the set of nodes, and $E_B$ is the set of weighted edges.
Every node $n_i \in V_B$ represents a profile $p_i \in P$, while every
edge $e_{i,j}$ represents a comparison
$c_{i,j} \in B \subseteq P \times P$. A schema-agnostic \textit{weighting
function} is employed to weight the edges, leveraging the co-occurrence patterns
of profiles in $B$: each edge is assigned a weight that is derived exclusively
from the (characteristics of the) blocks its adjacent profiles have in common.
For example, the \textbf{\textsf{ARCS}} function sums the inverse cardinality of
common blocks, assigning higher scores to pairs of profiles sharing smaller (i.e., more distinctive) blocks:
$ARCS(p_i, p_j, B)=\sum_{b_k \in B_i \cap B_j}1 / \|b_k\|$.
Similarly, all other weighting functions \cite{DBLP:conf/edbt/0001PPK16,
DBLP:journals/pvldb/SimoniniBJ16} assign high weights to edges connecting
profiles with strong co-occur\-rence patterns and low weights to 
% edges indicating 
casual co-occur\-ren\-ces. 

\begin{ex}
Figure \ref{fig:example}a shows a set of entity profiles, $P$. Figure
\ref{fig:example}b illustrates the block collection that is generated by
applying Token Blocking to $P$, i.e., by creating a separate block for every
token that appears in any attribute value of the input profiles (these
tokens are called \textbf{attribute value tokens} in the following).
Figure \ref{fig:example}c depicts the Blocking Graph
that is derived from the blocks of Figure \ref{fig:example}b,
when using the \textsf{ARCS} function for edge weighting.
\end{ex}

%\vspace{5pt}
%\noindent\textit{\hl{Shortcomings of the existing methods exploiting the Blocking Graph:}}\\
%\note{Giovanni: Maybe the following part can go elsewhere.}\\
%\textit{Challenges} |
Note that materializing and sorting all edges of a blocking graph is impractical
for large datasets, due to the resulting huge graph size (i.e., the number of
edges it contains)
 %Blocking Graph 
 \cite{DBLP:conf/edbt/0001PPK16}.
For this reason, all existing Meta-blocking methods
\cite{DBLP:conf/edbt/0001PPK16,
DBLP:journals/pvldb/SimoniniBJ16} discard low-weighted edges through a
\textit{pruning algorithm}, while building the Blocking Graph.
As a result, they retain only the most promising comparisons, which are
collected and employed for \textsf{Batch ER}.
% (i.e., without global sorting of the edges).
% In particular, two are the pruning algorithm relevant to our work: Cardinality
% Edge Pruning (CEP) and (2) Cardinality Node Pruning (CNP) retains the locally
% best comparisons, i.e., up to k top-weighted edges per node.
% CEP extracts the K top-weighted edges from the entire blocking graph, i.e.,
% the K globally best comparisons; then, the top-K comparisons are executed in
% batch.
% \note{CEP produce all the comparisons, then sort them: \textbf{unpractical}
% with large datasets} For the progressive ER problem, the edges of the Blocking
% Graph could be gathered and sorted by descending weights; however, this
Based on such a Blocking Graph, we present in Section \ref{sec:bg-methods} a
novel algorithm that generates comparisons in a progressive way.

\vspace{2pt}
\noindent\textbf{Neighbor List} | The \textit{Neighbor List} is the core data
structure of Sorted Neighborhood \cite{DBLP:conf/sigmod/HernandezS95} and its
derived methods (i.e., \textsf{PSN} \cite{DBLP:journals/tkde/PapenbrockHN15,
DBLP:journals/tkde/WhangMG13}).
% \textsf{schema-agnostic Sorted 
% Neighborhood}~\cite{DBLP:journals/pvldb/0001APK15},
It is a \textit{list of profiles} that is generated by sorting all profiles
alphabetically, according to the blocking keys that represent them.
This data structure is exploited to generate comparisons under the assumption
that the matching likelihood of any two profiles is analogous to their
\textit{proximity} after sorting. %in the Neighbor List.
% The Neighbor List can be built either from \textit{schema-based} blocking keys
% of from \textit{schema-agnostic} blocking keys.
% Figure \ref{fig:neighborlists}a show an example of \textit{schema-based}
% Neighbor List, derived sorting alphabetically the the blocking keys derived by
% concatenating the surname with the first two letter of the name; while, in
% \begin{ex} Figure \ref{fig:example}e shows an example of schema-agnostic
% Neighbor List, generated from the sorted list of blocking keys (the tokens) of
% the profiles depicted in Figure \ref{fig:example}a.
% \end{ex}

% In the literature,
The Neighbor List can be built 
%either 
from \textit{schema-based} \cite{DBLP:journals/tkde/Christen12}
%blocking keys
or from \textit{schema-agnostic} \cite{DBLP:journals/pvldb/0001APK15} blocking keys and is
typically employed to generate blocks:
% , both for batch (\cite{DBLP:conf/sigmod/HernandezS95}) and for progressive
% (\cite{DBLP:journals/tkde/PapenbrockHN15, DBLP:journals/tkde/WhangMG13})
% \textsf{ER}:
a window slides over the Neighbor List, and blocks correspond to groups of
profiles that fall into the same window. The size of the window is iteratively
incremented. The resulting blocks are called \textbf{redundancy-neutral blocks},
because the similarity of two profiles is not related to the number of
blocks they share;
% d by two profiles is not necessarily proportional to their, because 
the corresponding blocking keys might be close when sorted alphabetically, but rather dissimilar.
% As an example, consider the sorted keys in {Figure \ref{fig:example}d}:
% \texttt{`carl'} and \texttt{`ellen'} are placed in consecutive positions, but
% the corresponding profiles have nothing in common.

\begin{ex}
To understand the notion of redundancy-neutral blocks, consider
the sorted schema-agnostic blocking keys (i.e., the
attribute value tokens) of the profiles in Figure \ref{fig:example}a, which are
depicted in Figure \ref{fig:example}d. The keys \texttt{`carl'} and
\texttt{`ellen'} are placed in consecutive positions, but the corresponding
profiles have nothing in common. Figure \ref{fig:example}e shows the Neighbor
List that corresponds to this sorted list of schema-agnostic blocking keys.
% (i.e., the attribute value tokens) of the profiles in Figure \ref{fig:example}a.
\end{ex}

%\textit{Challenges } | In 
Note that in the schema-agnostic Neighbor List, every
profile typically has \textbf{multiple placements} (e.g., once for each attribute value
token) \cite{DBLP:journals/pvldb/0001APK15}. Hence, multiple distances can be measured for any
pair of matching profiles.
%in the Neighbor List 
In Section \ref{sec:sn-methods}, we present two approaches that leverage this
phenomenon to improve the early quality of \textsf{Progressive ER}.

\vspace{-5pt}
\section{Na\"ive Methods}
\label{sec:naiveSolutions}

\begin{figure}[t!]
	\centering
	\begin{tabular}{l}
		\includegraphics[width=0.47\textwidth]{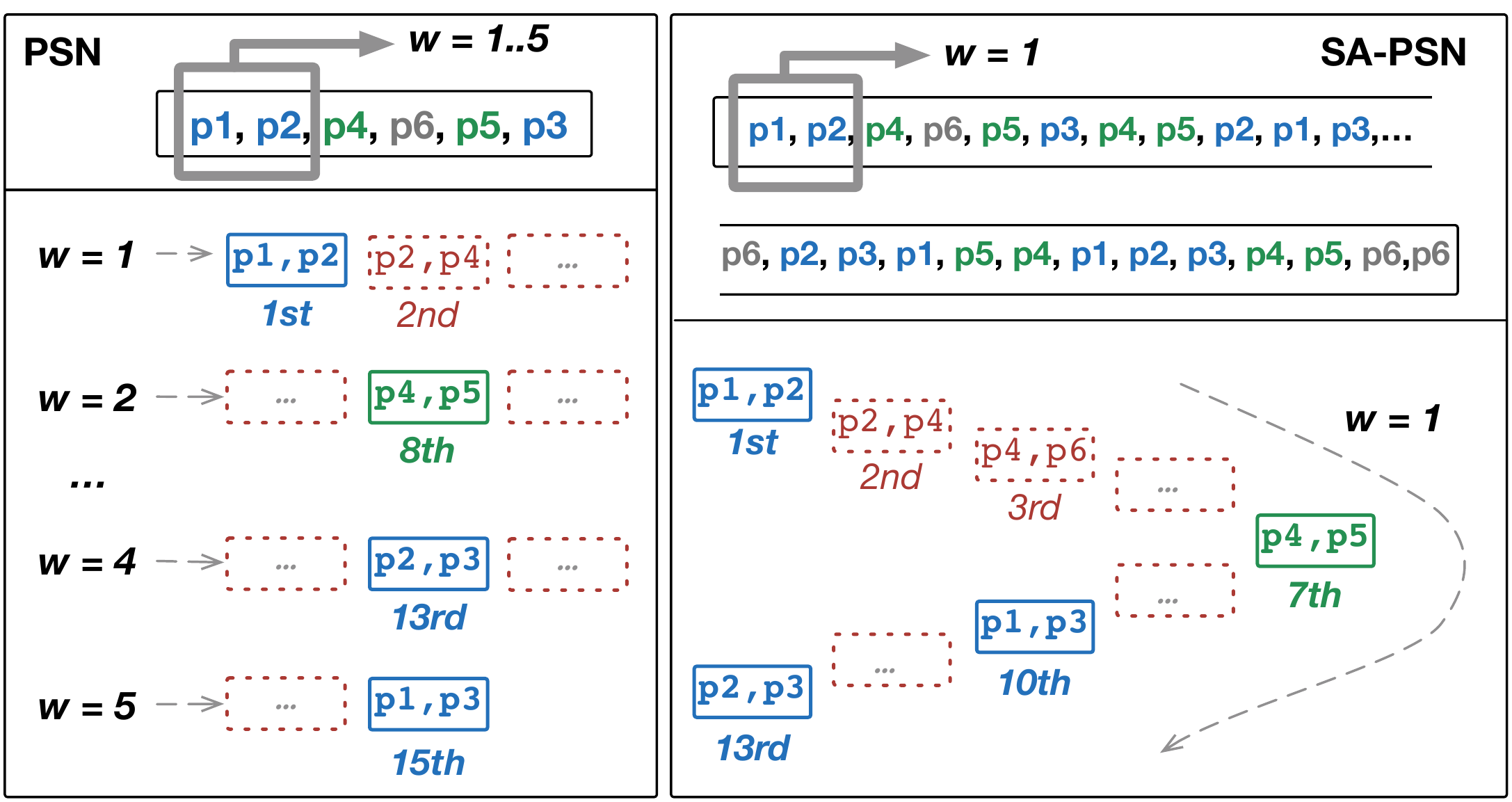}\\
		\hspace{0.7in}(a)              \hspace{1.45in}                     (b)
	\end{tabular}
	\vspace{-12pt}
	\caption{{\small Progressive emission of comparisons for (a) \textsf{PSN}, and
	(b) \textsf{SA-PSN}; dashed boxes indicate non-matching comparisons.}}
	\vspace{-12pt}
	\label{fig:sapsn}
\end{figure} 

Schema-based progressive methods (see Section \ref{sec:relatedWork}) are hard to
apply in a domain like Web data, where Variety renders the selection of
schema-based blocking keys into a non-trivial task.
Yet, we can convert the state-of-the-art schema-based progressive method
(\textsf{PSN}) into a schema-agnostic one with minor modifications,
 %(called \textsf{SA-PSN}), 
as explained in Section \ref{sec:sapsn}.
We can also adapt the established batch blocking method
(\textsf{Suffix Array
Blocking}~\cite{DBLP:conf/wiri/AizawaO05,DBLP:journals/tkde/Christen12,DBLP:journals/pvldb/0001APK15})
into a progressive method, based on the ideas of \textsf{HLR}
\cite{DBLP:conf/icde/AltowimM17,DBLP:journals/tkde/WhangMG13}, as explained in
Section \ref{sec:sapsab}.
%in the rest of this section.
However, our experimental analysis (Section~\ref{sec:experiments}) shows that
both methods have inherent limitations that lead to poor performance, thus
calling for the development of more advanced schema-agnostic progressive methods.

\vspace{-5pt}
\subsection{Schema-Agnostic PSN (\textbf{\textsf{SA-PSN}})}
\label{sec:sapsn}

The gist of this approach is to combine the sliding window with incremental
size of \textsf{PSN} \cite{DBLP:journals/tkde/WhangMG13} with the Neighbor List of the
schema-agnostic Sorted Neighborhood~\cite{DBLP:journals/pvldb/0001APK15}.
% every key corresponds to an \textit{attribute value token}, i.e., a token that
% appears in any attribute value of the profiles.
% \note{Giovanni: here I will insert an example and show both PSN and SA-PSN}.
The resulting method 
is called \textit{Schema-Agnostic Progressive Sorted
Neighborhood} (\textsf{SA-PSN}).

Inevitably, the Neighbor List of \textsf{SA-PSN} may involve consecutive places 
with the same profile (i.e., a profile which contains two alphabetically
consecutive tokens), or two profiles from the same source. The same applies to
entire windows.
% To combine the two methods, we simply apply the sliding window with increasing
% size of the former to the Profile List of the latter.
% profiles produced by schema-agnostic Sorted Neighborhood.
For this reason, the comparisons extracted from every window should involve
different profiles (\textsf{Dirty ER}), or profiles stemming from different
sources (\textsf{Clean-clean ER}).

%\begin{figure}[t!]\centering
%	\includegraphics[width=0.49\textwidth]{images/sapsn}
%	\vspace{-10pt}
%	\caption{PSN (a) and SA-PSN (b) progressive emission of comparisons; dashed boxes are non-matching comparisons.}
%	\vspace{-15pt}
%	\label{fig:sapsn}
%\end{figure}

%\note{Giovanni: I would insert the example for SA-PSN here}.
%The state-of-the-art progressive ER method PSN \cite{Wang., Pepe} associates every profile with a blocking key derived by a highly characteristic attribute.

\begin{ex}
Figure \ref{fig:sapsn} applies \textsf{PSN} and \textsf{SA-PSN} to 
the profiles of Figure \ref{fig:example}a.
For \textsf{PSN}, we assume that the schema of $p_1$ and $p_4$ describes all
other profiles,
% of the dataset in Figure \ref{fig:example}a 
%are described by , 
even $p5$ and $p6$, which represent unstructured data and would require an
\textit{information extraction} preprocessing step.
This assumption allows for defining a schema-based blocking key that
concatenates the surname and the first two letters of the name.
In this context, \textsf{PSN} in Figure \ref{fig:sapsn}a starts by emitting all
comparisons produced by the initial window size, $w=1$; then, it continues with
those comparisons entailed by window $w=2$ etc.
The final pair of matches is emitted during the $15^{th}$ comparison, i.e.,
after raising the window size to $w=5$.
In Figure \ref{fig:sapsn}b, \textsf{SA-PSN} applies the same procedure
to the schema-agnostic Neighbor List, finding all matching 
profiles within the initial window frame 
% of size one (
$w=1$, after the $14^{th}$ comparison. 
% Compared to \textsf{SA-PSN}, \textsf{PSN} entails significantly fewer
% comparisons per window (i.e., for each $w \in \{1..|P|\}$), but it may require a
% higher number of iterations (5 for \textsf{PSN} and 1 for \textsf{SA-PSN} in the
% example).
% depending on the blocking criterion that has been devised.
% Figure \ref{fig:sapsn}b also shows examples of: (i) repeated
% comparisons, e.g., $c_{12}$ is emitted as the $1^{st}$ and the $9^{th}$
% comparison within the same window frame, $w=1$; and (ii) coincidental
% proximity, since all 6 profiles are associated with the token \texttt{white}
% % (i.e., all the profile in the example), which 
% and are placed in random order at the end of the Neighbor List.
\end{ex}

The main advantage of \textsf{SA-PSN} is that it involves a parameter-free
functionality that requires no schema-based blocking key definition and has low
space and time complexities (see Section~\ref{sec:complexities}).
% Its space complexity is actually linear with respect to the size of the input
% dataset, $|P|$, i.e., $O($$\kpp$$\cdot$$|P|)$, because it merely keeps in
% memory the Neighbor List. Its time complexity is
% dominated by the sorting of blocking keys in alphabetical order, 
% $O($$\kpp$$\cdot$$|P|$$\cdot$$\log$$(\kpp$$\cdot$$|P|))$, thus ensuring high
% scalability.
On the flip side, \textsf{SA-PSN} may perform
% (1) \textit{coincidental proximity}: the proximity of two profiles in the list
% may coincidental; \note{example}\\
% (1) \textit{coincidental ordering}: the order in which comparisons are
% performed is random to a large extent; \note{example}; (Notice that PSN
% suffers of coincidental ordering as well)
\textbf{\textit{repeated comparisons}}: the same pair of profiles might co-occur
multiple times in the various windows. For example, in Figure \ref{fig:sapsn}b,
$c_{12}$ is emitted as the $1^{st}$ and the $9^{th}$
comparison within the same window frame, $w=1$.
Moreover, the proximity of two profiles in the list may be partially random;
if more than two profiles share the same blocking key, they are inserted with a
relatively random order in the Neighbor List. 
We call this phenomenon \textbf{\textit{coincidental proximity}}. As an
example, consider all 6 profiles in Figure \ref{fig:sapsn}b that are
associated with the token \texttt{white};
% (i.e., all the profile in the example), which 
they are placed in random order at the end of the Neighbor List.
Note that \textsf{PSN} also suffers from coincidental proximity, which
is a critical point to consider when devising the schema-based blocking keys.

\vspace{-12pt}
\subsection{Schema-Agnostic Progressive SAB
(\textsf{SA-PSAB})}
\begin{figure}[t!]\centering
	\includegraphics[width=0.2\textwidth]{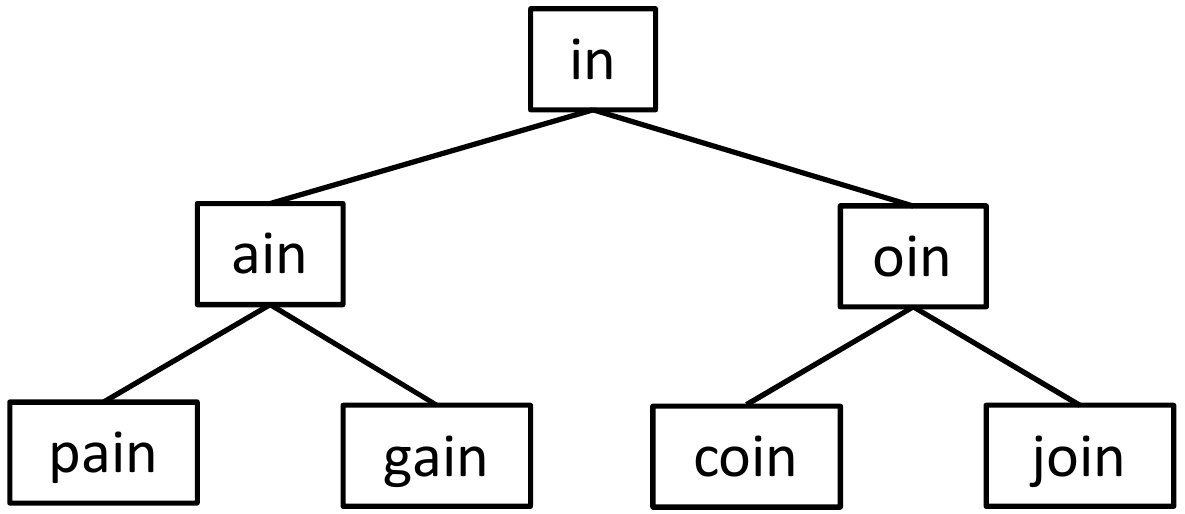}
	\vspace{-8pt}
	\caption{An example of a suffix tree employed by \textsf{SA-PSAB}.}
	\vspace{-12pt}
	\label{fig:sTree}
\end{figure}

%\vspace{-4pt}
\label{sec:sapsab}
Suffix Arrays Blocking (\textsf{SAB}) is a schema-based blocking technique that
addresses noise at the start of blocking keys by converting them into all
suffixes that contain at least $l_{min}$ characters (\textit{minimum suffix
length}) \cite{DBLP:conf/wiri/AizawaO05,DBLP:journals/tkde/Christen12}.
Basically, \textsf{SAB} uses these suffixes to generate hierarchical
blocks such that: the lowest levels in the hierarchy correspond to blocks
generated with the initial blocking keys (e.g., ``coin'', ``join'', ``pain'',
''gain''), the intermediate levels correspond to blocks generated with the
intermediate suffixes (e.g., ``oin'', ``ain''), and the highest levels
correspond to blocks generated with the shortest allowed suffixes (e.g., ``in''
for $l_{min}$=2).
This hierarchy of blocks follows the corresponding hierarchies of suffixes,
which we call \textit{suffix forest}.
Notice that there are as many \textit{suffix trees} as the number of distinct
suffixes of size $l_{min}$.
Figure~\ref{fig:sTree} depicts an example of suffix tree.
% Hence, \textsf{SAB} ensures that two matching profiles with different blocking
% keys might still co-occur in as many blocks as the number of key suffixes they
% share.
% To avoid an excessive growth in blocks and comparisons, \textsf{SAB} sets an
% upper limit on the number of profiles per block through the parameter
% \textit{maximum block size} ($b_{max}$).
To address the Variety of Big Data in a schema-agnostic fashion, every
attribute value token can be considered as a blocking key
\cite{DBLP:journals/pvldb/0001APK15}.%, \textsf{SAB}.

%To adapt schema-agnostic \textsf{SAB} to \textsf{Progressive ER}, we disregard the second configuration parameter, namely $b_{max}$.
%Thus, every suffix of minimum length (e.g., ``$in$'' for $l_{min}$=2) creates a \textit{suffix tree} with that suffix at its root.
%The children of the root correspond to suffixes with one additional character (e.g., ``$oin$'' and ``$ain$'').
%The node labels are incremented in the same fashion until we reach the tree leaves, which correspond to the largest attribute value tokens that end with the
%root token (e.g., ``$coin$'' and ``$pain$'').
%The resulting tree for our example appears in Figure~\ref{fig:sTree}. 

%Note that every suffix tree forms a hierarchy of blocks, where every profile $p_i$ is
%placed at the lowest level of  every tree, whose root label is contained in the
%suffixes of $p_i$'s attribute value tokens.
%Overall, our modification yields a group of such trees, which we call
%\textit{suffix forest}.

The processing of an individual suffix tree follows a ``\textit{leaves first, root last}'' approach.This means that the candidate pairs that appear in a lower level
block (e.g., in a ``leaf block'') are emitted before candidate pairs in a higher level block (e.g., a ``root block'').
% \note{Now the Related Work is at the end; change the reference to HRP.}
Thus, for the entire suffix forest, the processing starts from the leaf node of the
lowest layer (i.e., the overall largest attribute value token) and moves on to
the tree roots; nodes of the same layer are ordered in increasing number of
comparisons (i.e., the smallest nodes are processed first).
We call the resulting method \textit{Schema-Agnostic Progressive Suffix Arrays Blocking}
(\textbf{\textsf{SA-PSAB}}).

Despite its complex functionality, \textsf{SA-PSAB} is probably the
easiest-to-configure \textsf{HRP} or \textsf{OLR} progressive method. Its
data-driven functionality 
%renders unnecessary the $L$+2 parameters of \textsf{HRP}, due to .
%It 
simply extracts from every profile all attribute value tokens and 
for every such token, it considers as blocking keys all suffixes with at least
$l_{min}$ characters. Therefore, $l_{min}$ is the
only configuration parameter of \textsf{SA-PSAB}.
Note also that \textsf{SA-PSAB} can be considered as the
schema-agnostic version of the hierarchical progressive method proposed
in~\cite{DBLP:conf/icde/AltowimM17}.

% Regarding the complexity of \textsf{SA-PSAB}, its initialization phase is
% dominated by the sorting of all suffixes (i.e., tree nodes) in non-increasing
% order of length and non-decreasing order of comparisons, i.e.,
% $O($$\bar{s_e}$$\cdot$$|P|$$\cdot$$\log$$(\bar{s_e}$$\cdot$$|P|))$,
% where $\bar{s_e}$ is the average number of suffixes per profile. Its emission
% phase simply returns the next comparison in a tree node, thus exhibiting a
% constant time complexity, $O(1)$. Finally, the space required to the suffix
% forest in memory is $O($$\bar{s_e}$$\cdot$$|P|)$.

% See Appendix \ref{sec:complexities} for more details about its space and time
% complexity.

\vspace{-10pt}
%\vspace{-6pt}
\section{Advanced Methods}
\label{sec:approach}

We now introduce more elaborate methods for schema-agnostic
\textsf{Progressive ER}, using a broad spectrum of techniques. 
%A detailed complexity analysis for all methods is provided in the Appendix.
We distinguish them into two categories: the \textit{similarity-based} ones
(Section~\ref{sec:sn-methods}), which employ a \textit{weighted} Neighbor List,
and  the \textit{equality-based} ones (Section~\ref{sec:bg-methods}), which
employ a Blocking Graph.
The former are based on the \textit{similarity principle}, the
latter on the \textit{equality principle} (see Section~\ref{sec:introduction}).
% They are presented in Sections~\ref{sec:sn-methods} and~\ref{sec:bg-methods},
% respectively. 
% The former achieve the highest performance over structured (relational) data,
% while the latter excel over the semi-structured or unstructured Web data
% (Section~\ref{sec:experiments}).

Note that all our methods employ a data structure called
\textbf{Comparison List}, which essentially constitutes a list of comparisons
sorted in non-increasing order of matching likelihood. Its purpose is to store
the best comparisons that were detected during the initialization phase so that
they are efficiently emitted during the emission phase. Whenever the Comparison
List gets empty, it is refilled with the next batch of the best remaining
comparisons, during the next 
% call of the
emission phase.

\vspace{-5pt}
\subsection{Similarity-based Methods}
\label{sec:sn-methods}

These methods extend the similarity principle of
\textsf{SA-PSN}, assuming that the closer the blocking keys of two profiles are,
when sorted alphabetically, the more likely they are to be matching.
%In the previous section, we explained that 
As explained above,
\textsf{SA-PSN}
suffers from two drawbacks:
it contains numerous repeated comparisons and it defines a processing order of comparisons that is
partially random, due to \textit{coincidental proximity}. 
% (see Section \ref{sec:sapsn}).
To address both disadvantages, we propose the use of a \textbf{weighted
Neighbor List}, which employs a \textit{weighting scheme} in order to associate
every comparison with a numerical estimation of the likelihood that it involves a pair of
matching  profiles.
This weighting scheme leverages the Neighbor List, with a functionality that is both schema- and domain-agnostic.
Consequently, our approach addresses inherently the Variety of Web data.

% We should stress at this point that
% Meta-blocking~\cite{DBLP:conf/edbt/0001PPK16} cannot be used in this case: it does not apply to the Neighbor List of \textsf{SA-PSN}, since it requires redundancy-positive blocks as input, whereas \textsf{SA-PSN} produces redundancy-neutral blocks.
% Therefore, \textsf{SA-PSN} calls for a novel weight-based functionality.

To this end, we propose  the \textit{Relative Co-occurrence Frequency}
(\textbf{\textsf{RCF}}) weighting scheme. \textsf{RCF} counts how many times
a pair of profiles lies at a distance of $w$ positions in the Neighbor List and then 
normalizes it  by the number of positions corresponding to each profile.
To efficiently implement \textsf{RCF} and
\textit{weighted} Neighbor List, we go beyond Neighbor List by introducing a
new data structure called \textbf{Position Index}. In essence, this is an
inverted index that associates every profile (id) with its positions in the
Neighbor List. Thus, it is generic enough to accommodate any weighting scheme
that similarly to \textsf{RCF} relies on the co-occurrence frequency of profile
pairs.

Below, we present two algorithms that exploit the \textsf{RCF} weighting scheme.
Both of them are compatible with any other schema-agnostic weighting scheme that
infers the similarity of profiles exclusively from their co-occurrences in the
incremental sliding window.
The core idea of these algorithms is to trade a higher computational cost of the
initialization phase, and probably the emission phase, for a significantly
better comparison order. 
%Note that both of them are compatible with any other
%weighting scheme, except for \textsf{RCF}.

%\vspace{5pt}
\subsubsection{Local Schema-Agnostic PSN (\textbf{\textsf{LS-PSN}})}
\label{sec:lspsn}
\begin{figure}[t!]\centering
	\includegraphics[width=0.47\textwidth]{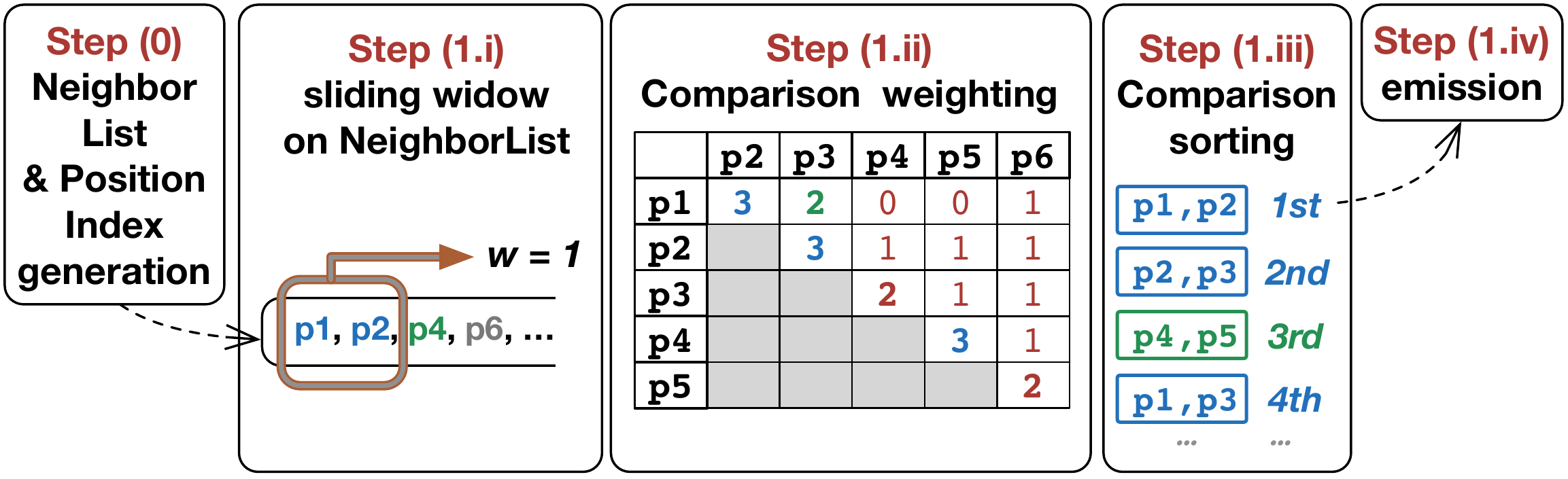}
	\vspace{-8pt}
	\caption{Applying \textsf{LS-PSN} to the profiles of Figure
	\ref{fig:example}a.}
	\vspace{-8pt}
	\label{fig:ex_psn}
\end{figure}

This approach applies the selected weighting scheme only to
the comparisons of a specific window size, thus defining a local execution
order. At its core lie two data structures:

%|\begin{enumerate}[i)]
%[leftmargin=*]
%\item 
\emph{i)} $NL$, which is an array that encapsulates the Neighbor List  such that
$NL[i]$ denotes the profile id that is placed in the $i^{th}$ position of the Neighbor List.
An exemplary $NL$ array is shown in Step 1.i of Figure \ref{fig:ex_psn}.

% \item 
\emph{ii)} 
$PI$, which stands for \textit{Position Index}, is an inverted index that
points from profile ids to positions in $NL$.
It is implemented with an array that uses profile ids as indexes, such that
$PI[i]$ returns the list of the positions associated with profile $p_i$ in $NL$.
This array accelerates the estimation of comparison weights, since it minimizes
the computational cost of retrieving the neighbors of any profile in the current
window, as described below. Note that instead of a Position Index,
\textsf{LS-PSN} could use a hash index that has comparisons as keys and weights
as values. This approach, however, would increase both the space and the time
complexity of comparison weighting.
%\end{enumerate}

\begin{algorithm}[t]
{\scriptsize
  \LinesNumbered %
  \SetAlgoLined %SetLine
  \SetAlgoVlined %SetVline 
  \KwIn{(i)~Profile~collection~$P$,~(ii)~Weighting~scheme,~$wScheme$}
  \KwOut{The overall best comparison}
  $windowSize$ = 1\;
  $ComparisonList$ $\leftarrow$ $\emptyset$\;
  $NL$[] $\leftarrow$ buildNeighborList($P$)\;
  $PI$[] $\leftarrow$ buildPositionIndex($NL$[])\;
  \BlankLine
  \ForEach{$p_i$ $\in$ $P$}{
  	$distinctNeighbors$ $\leftarrow$ $\emptyset$; // a set containing distinct
  	neighbors\;
  	$frequency$[] $\leftarrow$ $\emptyset$\;
  	\ForEach{$position$ $\in$ $PI$[$i$]}{
  		$p_j$ $\leftarrow$ $NL$[$position$+$windowSize$]\;
  		\If {isValidNeighbor($p_j$)} {
  			$frequency$[$j$]++\;	
  			$distinctNeighbors$.add($j$)\;
  		}
  		
  		$p_k$ $\leftarrow$ $NL$[$position$-$windowSize$]\;
  		\If {isValidNeighbor($p_k$)} {
  			$frequency$[$k$]++\;	
  			$distinctNeighbors$.add($k$)\;
  		}
  	}
  	
  	\ForEach{$j$ $\in$ $distinctNeighbors$} {
  		$weight_{i,j}$ $\leftarrow$ $wScheme$($frequency$[$j$], $j$, $i$)\;
  		$ComparisonList$.add(getComparison($i$, $j$, $weight_{i,j}$)\;
	}
  }
  
  sortInDescreasingWeight($ComparisonList$)\;
	
  \KwRet{$ComparisonList$.removeFirst()}\;
  \caption{{\small Initialization phase for \textsf{LS-PSN}.}}
  \label{alg:initLSPSN}
}
\end{algorithm}
\begin{algorithm}[t]
{\scriptsize
  \LinesNumbered %
  \SetAlgoLined %SetLine
  \SetAlgoVlined %SetVline 
  \KwOut{The next best comparison}
  \While {$ComparisonList$.isEmpty()} {
  	$windowSize$++\;
  	  \If {$NL$.size() $<$ $windowSize$ } {
  		return null\;	
  		}
  	\tcc{repeat lines 5 - 20 in Algorithm \ref{alg:initLSPSN}}
  }
  \KwRet{$ComparisonList$.removeFirst()}\;
  \caption{{\small Emission phase for \textsf{LS-PSN}.}}
  \label{alg:emitLSPSN}
}
\end{algorithm}

Based on these data structures, the initialization phase of \textsf{LS-PSN} is
outlined in  Algorithm \ref{alg:initLSPSN}.
Initially, it sets the window size to 1 (Line 1), considering only consecutive
profiles. Then, it creates its data structures (Lines 2-4) and for
every profile $p_i$ (Line 5), it iterates over all its positions in the Position
Index (Line 8). In every position, \textsf{LS-PSN} checks the neighbors
in both directions, i.e., the profiles located $windowSize$ places before and after
$p_i$ (Lines 13 and 9, respectively) - provided that the corresponding positions
are within the limits of the Neighbor List.
For every neighbor $p_j$, \textsf{LS-PSN} checks if $j{<}i$ (Line 10) and $k{<}i$ (Line 14) to avoid
repeated comparisons.
For every valid neighbor, \textsf{LS-PSN} increases its
frequency (Lines 11 and 15) and adds it into the set of neighbors (Lines 12 and 16).
Then, the overall weight for every comparison is computed
according to the selected weighting scheme (Line 18) - assuming a
comparison between $p_i$ and $p_j$, i.e., $c_{i,j}$, the corresponding
\textsf{RCF} weight is equal to $\frac{frequency[j]}{PI[i].length()+PI[j].length()-frequency[j]}$. 
Finally, all comparisons are aggregated and sorted from the highest weight to
the lowest (Line 20) and the top one is returned (Line 21).

Note that Algorithm \ref{alg:initLSPSN} pertains to \textsf{Dirty ER}. Yet, it
can be adapted to \textsf{Clean-clean ER} with two minor modifications:
\emph{(i)} Line 5 iterates over the profiles of $P_1$, and \emph{(ii)} in Lines
10 and 14, a neighbor $p_j$ is considered valid only if $p_j \in P_2$.

The emission phase of \textsf{LS-PSN} is illustrated in Algorithm
\ref{alg:emitLSPSN} and is common for both \textsf{Clean-clean} and
\textsf{Dirty ER}:
if the Comparison List corresponding to the current window is not empty, the top weighted one is removed and returned as output (Line 3). 
If the list is empty, the window size is incremented (Line 2) and the process for
extracting all comparisons of the new window (Lines 5 - 20 in Algorithm
\ref{alg:initLSPSN}) is repeated.
After each emission, the processing can be interrupted. In the
worst case, the emission phase is terminated when the window size is equal to
the size of the Neighbor List (Lines 3-4). This means that the window is so
large that it ends up comparing every profile with all others.

\begin{ex}
We demonstrate the functionality of \textsf{LS-PSN} by applying it to
the profiles of Figure~\ref{fig:example}a. 
The result appears in Figure~\ref{fig:ex_psn}. Step 0 extracts all blocking keys and sorts them
alphabetically, while Step 1.i forms $NL$ and slides a window of size 1
over it. In Step 1.ii, we see the result of the nested loops in Lines 5 - 16
for the \textsf{RCF} weighting scheme for $windowSize$=1. In Step
1.iii, all comparisons are weighted and sorted from the highest to the lowest
weight. Finally, the sorted comparisons are emitted one by one in Step 1.iv.
Note that the first three comparisons correspond to the three pairs of duplicate profiles.
\end{ex}

% \textbf{Complexity Analysis.} \textsf{LS-PSN} mainly keeps in memory the
% Neighbor List along with the Position Index. For both data structures, the space
% complexity is $O($$\kpp$$\cdot$$|P|)$, depending linearly on the number of input
% profiles. Similar to \textsf{SA-PSN}, the time complexity of the
% initialization phase is dominated by the sorting of blocking keys in
% alphabetical order, i.e.,
% $O($$\kpp$$\cdot$$|P|$$\cdot$$\log$$(\kpp$$\cdot$$|P|))$. Finally, the time
% complexity of the emission phase is usually constant, $O(1)$, simply emitting
% the next comparison from the Comparison List. Whenever this list gets empty,
% \textsf{LS-PSN} renews its contents by repeating their initialization phase,
% raising the complexity to  
% $O($$\kpp$$\cdot$$|P|$$\cdot$$\log$$(\kpp$$\cdot$$|P|))$; the only difference
% with the initialization phase is the incremented window size.

%\vspace{5pt}
\subsubsection{Global Schema-Agnostic PSN (\textbf{\textsf{GS-PSN}})} 
\label{sec:gspsn}
% The {\color{red}main drawback of \textsf{LS-PSN} is the unstable response time of
% its emission phase. Its time complexity is usually constant, $O(1)$, but
% whenever the window size gets incremented, the entire process of weighting and
% storing all comparisons for the new window size is repeated, increasing the time
% complexity to $O($$\bar{|p|}$$\cdot$$|P|)$. \textsf{GS-PSN} aims to overcome
% this drawback by offering an emission phase of consistently constant time
% complexity.}
{
%\color{blue}
The main drawback of \textsf{LS-PSN} is the \textit{local} execution order it
defines for a specific window size. This means that \textsf{LS-PSN} is likely to emit
the same comparison(s) multiple times, for two or more different window
sizes, since it does not remember past emissions.
\textsf{GS-PSN} aims to overcome this drawback by 
defining a \textit{global} execution order for all the comparisons in
a range of window sizes $[1, w_{max}]$.
% Recall that \textsf{LS-PSN} defines a \textit{local} execution order for a specific window size.
}
%Unlike \textsf{LS-PSN}, which defines a \textit{local} execution order for a specific window size, 
%\textsf{GS-PSN} defines a \textit{global} execution order for all comparisons in a range of window sizes $[1, w_{max}]$.
To this end, its initialization phase differs from Algorithm \ref{alg:initLSPSN} in that
Line 1 is converted into an iteration over all window sizes in $[1, w_{max}]$;
this loop starts before Line 8 and ends before Line 20.
This allows for a simpler emission phase, which just returns the next best
comparison, until the Comparison List gets empty.

Compared to \textsf{LS-PSN}, \textsf{GS-PSN} takes into account more
co-occurrence patterns, 
%than \textsf{LS-PSN}, 
when determining comparison weights.
Consequently, its matching likelihood estimations are expected to be more
accurate than those of \textsf{LS-PSN}. This is achieved through an additional
%incorporating one more 
configuration parameter, $w_{max}$, 
%in order to 
which eliminates all repeated
comparisons in a particular range of windows. 
% As a result, \textsf{GS-PSN} occupies more space,
% $O(w_{max}{\cdot}\kpp{\cdot}|P|)$, since in the worst case, its Comparison List
% contains 1 comparison per position in the Neighbor List for every window size. 
% The time complexity of the initialization phase is the same as for
% \textsf{LS-PSN}, being dominated by the sorting of blocking keys in alphabetical
% order, while the emission phase exhibits a consistently constant time
% complexity, $O(1)$.
%\textsf{GS-PSN} also eliminates all repeated comparisons by computing in advance all comparisons associated with the window sizes it considers (in contrast, \textsf{LS-PSN} is likely to emit the same comparison for two or more different window sizes, since it does not remember past emissions).
%Note also that 

\vspace{-5pt}
\subsection{Equality-based Methods}
\label{sec:bg-methods}
\label{sec:cep}

These methods rely on the equality principle of 
%The methods in this category rely on a 
the redundancy-positive blocks 
that are derived from any 
%suitable 
schema-agnostic blocking method or
workflow~\cite{DBLP:conf/edbt/0001PPK16}: they assume that the more blocks two
entities share, the more likely they are to be matching. From these blocks, we
extract the Blocking Graph of Meta-blocking, using the weights of its edges as approximations for the matching likelihood of the
corresponding comparison.
In particular, we order the graph edges in decreasing weight in order to produce
a sorted list of comparisons at the level of individual blocks or profiles.
Below, we propose two novel algorithms of this type.

\subsubsection{Progressive Block Scheduling (\textbf{\textsf{PBS}})}
\label{sec:pbsMethod}

% In the following, we propose a new algorithm, \textit{Progressive Block
% Scheduling} (\textbf{\textsf{PBS}}), which 
This algorithm is specifically designed for \textsf{Progressive ER}, but relies
on a \textsf{Batch ER} technique.
%We note that in the  context of 
Indeed, \textit{Block Scheduling} has
been proposed in order to optimize the processing order of blocks
in the context of \textsf{Batch ER}, based on the probability that they
contain duplicates~\cite{DBLP:series/synthesis/2015Christophides}. It assigns to
every block a weight that is proportional to the likelihood that it contains
duplicates and then, it sorts all blocks in descending weight order.
Even though we would like to use such a functionality for \textsf{Progressive
ER}, it is not applicable, because: \emph{(i)} its weighting cannot generalize
to \textsf{Dirty ER}, applying exclusively to \textsf{Clean-clean ER}, and
\emph{(ii)} it does not specify the execution order of comparisons inside blocks
with more than two profiles.

Our algorithm, \textsf{PBS}, deals with both issues in two ways:

\emph{(1)} \textsf{PBS} introduces a weighting mechanism that applies uniformly
to \textsf{Clean-clean} and \textsf{Dirty ER}. In fact, it relies on the
reasonable hypothesis that the smaller a block is, the more distinctive information it encapsulates
and the more likely it is to contain duplicate profiles, and vice versa:
the larger a block is, the more frequent is the corresponding
blocking key/token and the more likely it corresponds to a stop word, thus
ingesting noise into the matching likelihood of two entities. Therefore,
our scheme sets weights inversely proportional to block
cardinalities (i.e., $1/\|b_i\|$) and sorts blocks in decreasing weights; the
fewer comparisons a block entails, the higher it is ranked.

\emph{(2)} \textsf{PBS} defines the processing order of comparisons inside
every block using the Blocking Graph. 
%This means that 
For each block $b_i$ with
$\|b_i\|$$>$$1$, \textsf{PBS} associates all comparisons with a weight derived
from any schema-agnostic weighting scheme of Meta-blocking. Then, it
sorts them from the highest weight to the lowest one. 

It is worth noting that all repeated comparisons are discarded before computing
their weight. In fact, the efficient detection of repeated comparisons is
crucial for \textsf{PBS}.
This functionality is based on a data structure called
\textbf{Profile Index},
which constitutes an inverted index that associates every profile with the ids of the
blocks that contain it. In this way, it facilitates the efficient
computation of comparison weights, similar to the Position Index of
\textsf{LS}/\textsf{GS-PSN}. Note that the Profile Index is generic enough to
accommodate any weighting scheme that is based on the block co-occurrence
frequency of profile pairs. 

In practice, the Profile Index is implemented as a two-dimensional array.
The first dimension is of size $|P|$ 
such that $ProfileIndex[i]$ points to an array that contains all ids of
the blocks involving profile $p_i$. As a result, the second dimension
contains arrays of variable length. The block ids in every such array are sorted
from the lowest to the highest one in order to ensure high efficiency for the
two operations that are built on top of the Profile Index.

\begin{algorithm}[t]
{\scriptsize
  \LinesNumbered %
  \SetAlgoLined %SetLine
  \SetAlgoVlined %SetVline 
  \KwIn{(i) Profile collection $P$, (ii)~Weighting~scheme,~$wScheme$}
  \KwOut{The overall best comparison}
%   $B$ $\leftarrow$ tokenBlocking($E$)\;
%   $B'$ $\leftarrow$ blockPurging($B$)\;
%   $B''$ $\leftarrow$ blockFiltering($B'$)\;
  $B$ $\leftarrow$ buildRedundancyPositiveBlocks($P$)\;
  $B'$ $\leftarrow$ blockScheduling($B$)\;
  $ProfileIndex$ $\leftarrow$ buildProfileIndex($B'$)\;
  $b_k$ $\leftarrow$ $B'$.removeFirst()\;
  $ComparisonList$	$\leftarrow$ $\emptyset$\;
  \ForEach{$c_{ij}$ $\in$ $b_k$}{
  	$B_i$ $\leftarrow$ $ProfileIndex$.getBlocks($p_i$)\;
  	$B_j$ $\leftarrow$ $ProfileIndex$.getBlocks($p_j$)\;
    \If {nonRepeated($k$, $B_i$, $B_j$)} {
    	$w_{i,j}$ $\leftarrow$ $wScheme$($k$, $B_i$, $B_j$)\;
    	$ComparisonList$.add(getComparison($i$, $j$, $w_{i,j}$))\;
  	}
  }
  sortInDescreasingWeight($ComparisonList$)\;
	
  \KwRet{$ComparisonList$.removeFirst()}\;
  \caption{{\small Initialization phase for \textsf{PBS}.}}
  \label{alg:initPBS}
}
\end{algorithm}
\begin{algorithm}[t]
{\scriptsize
  \LinesNumbered %
  \SetAlgoLined %SetLine
  \SetAlgoVlined %SetVline
  \KwOut{The next best comparison}
  \If {$ComparisonList$.isEmpty()} {
  	\tcc{repeat lines 4 - 12 in Algorithm \ref{alg:initPBS}}
  }
  \KwRet{$ComparisonList$.removeFirst()}\;
  \caption{{\small Emission phase for \textsf{PBS}.}}
  \label{alg:emitPBS}
}
\end{algorithm}

The first operation is the \textit{Least Common Block Index} (\textbf{\textsf{LeCoBI}})
condition, which checks whether a comparison is repeated in Line 9
of Alg. \ref{alg:initPBS}:
given a comparison $c_{ij}$ in block $b_Y$, the \textsf{LeCoBI} condition identifies the
least common block id, X, between the profiles $p_i$ and $p_j$ and compares it
with the id of $b_Y$, Y. If the two ids match ($X=Y$), $c_{ij}$ corresponds to
a new comparison. Otherwise $X<Y$, which means that $c_{ij}$ has already been
compared in block $b_X$, but is repeated in block $b_Y$. Note that $X>Y$ is
impossible, because the id of every block indicates its position in the
processing list after sorting all blocks in increasing cardinalities 
(i.e., $b_k$ denotes the block placed in the $k^{th}$ position after
sorting). Note also that by ordering the block ids of the second dimension in
increasing order, the Profile Index minimizes the checks required for detecting
the least common block id, thus accelerating the \textsf{LeCoBI} condition.

The second operation is \textit{Edge Weighting}, which in Line 10 of
Alg. \ref{alg:initPBS} infers the
matching likelihood of every comparison from the weight of the
corresponding edge in the blocking graph. 
Given a non-repeated comparison $c_{ij}$, it compares the block lists associated with profiles
$p_i$ and $p_j$ in order to estimate the number of blocks they share. 
This number, which lies at the core of practically all Meta-blocking weighting
schemes~\cite{DBLP:conf/edbt/0001PPK16}, can be derived from the evidence provided by the Profile Index. 
Note that by ordering the block ids of its second dimension in increasing order, the Profile Index allows for accelerating Edge Weighting by traversing the two block lists in parallel.
 
On the whole, the initialization phase of \textsf{PBS} appears in
Algorithm~\ref{alg:initPBS}. Initially, it creates a redundancy-positive block
collection and sorts its elements in non-decreasing order of comparisons (Lines
1-2). Then, it builds the corresponding Profile Index (Line 3) and goes on to
remove the first (i.e., smallest) block, iterating
over its comparisons (Lines 4-6). For every comparison $c_{ij}$, \textsf{PBS}
gets the block lists that are associated with profiles 
$p_i$ and $p_j$ from the Profile Index (Lines 7-8). Based on these lists, it
evaluates the \textsf{LeCoBI} condition, checking whether $c_{ij}$ is repeated or not (Line 9). If
$c_{ij}$ is a new comparison, it is placed in the Comparison List along with the 
weight of the corresponding Blocking Graph edge 
(Lines 10-11). 
After processing all comparisons in the current block, the elements of the Comparison List are
sorted in decreasing weight and the first one is emitted (Lines 12-13).

The emission phase of \textsf{PBS} appears in Algorithm \ref{alg:emitPBS}.
If the Comparison List is empty, it processes the next block $b'$$\in$$B'$,
applying the Lines 4-12 of Algorithm \ref{alg:initPBS} to it. Otherwise, the
next best comparison is emitted from the Comparison List.

\begin{figure}[t!]\centering
	\includegraphics[width=0.42\textwidth]{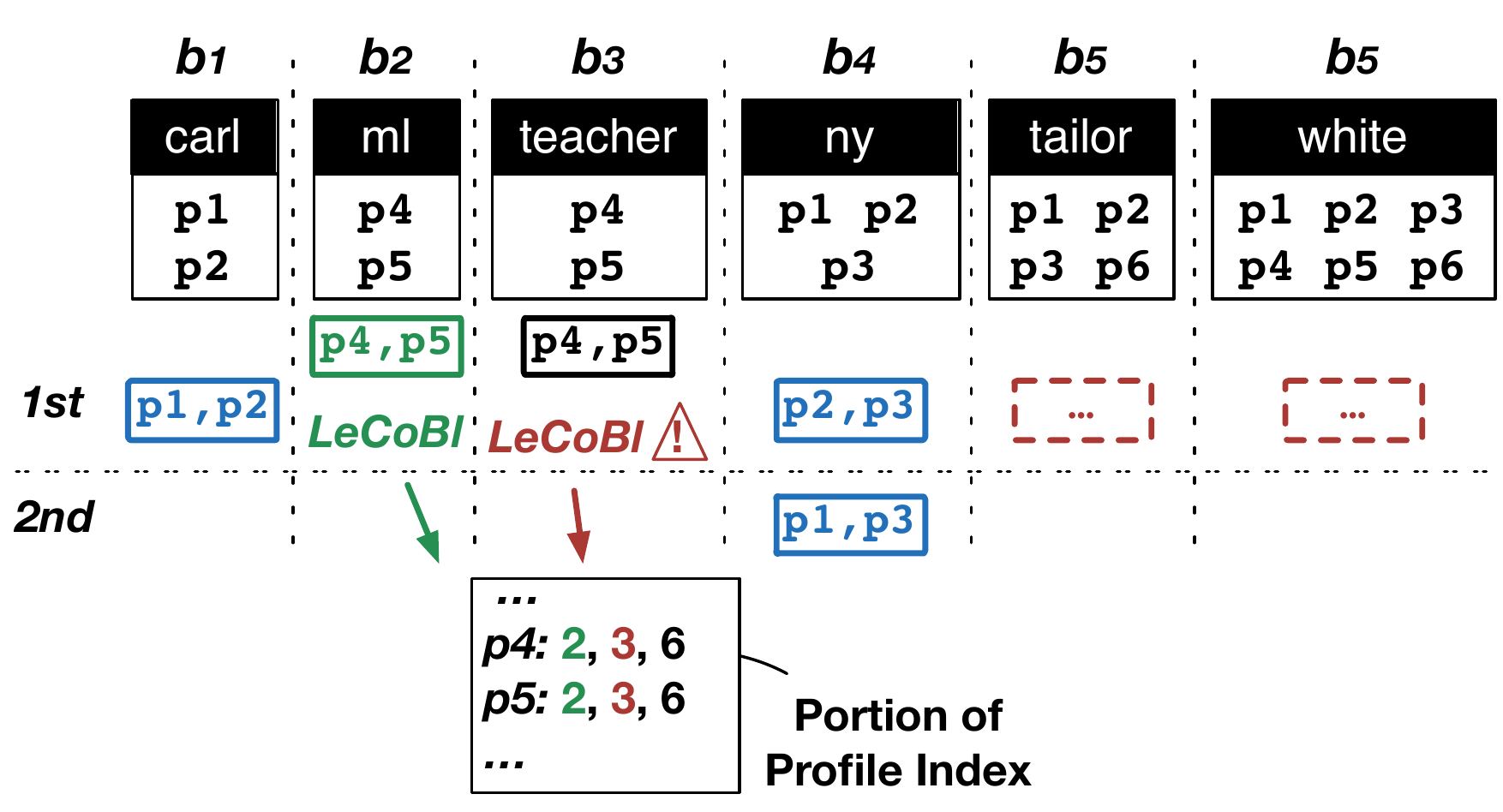}
	\vspace{-8pt}
	\caption{Applying \textsf{PBS} to the blocks of Figure
	\ref{fig:example}b.}
	\vspace{-14pt}
	\label{fig:pbs}
\end{figure}

%\vspace{-5pt}
\begin{ex}
Figure \ref{fig:pbs} illustrates the functionality of \textsf{PBS} by applying
it to the blocks of Figure \ref{fig:example}b. First, it sorts them in
non-decreasing cardinality and assigns to each one an
incremental block id that indicates its processing order
(note that we chose a random permutation of the blocks that have
the same number of comparisons, without affecting the end result).
Then, \textsf{PBS} processes the sorted list of blocks one block at a time, 
emitting iteratively the comparisons entailed in every block.
Inside every block, all comparisons that satisfy the \textsf{LeCoBl}
condition (i.e., non-repeated comparisons) are sorted according to the
corresponding edge weight in the Blocking Graph of Figure \ref{fig:example}c.
For instance, when \textsf{PBS} processes $b_2$,
the comparison $c_{45}$ satisfies the \textsf{LeCoBl} condition, since the least
common block id shared by $p4$ and $p5$ is 2.
This means that \textsf{PBS} encounters $c_{45}$ for the first time in $b_2$,
assigning the edge weight 1.33 to it.
In contrast, when \textsf{PBS} processes $b_3$, the comparison $c_{45}$ does not satisfy the \textsf{LeCoBl} condition anymore and is thus discarded.
\end{ex}
%\vspace{-5pt}
% \textbf{Complexity Analysis.} The space complexity of \textsf{PBS} is dominated
% by the space requirements of the Profile Index, i.e., $O($$\kpp$$\cdot$$|P|)$. The time complexity of its
% initialization time is dominated by the sorting of blocks in non-decreasing
% comparisons, i.e., $O(|B|{\cdot}\log|B|)$. In contrast, the cost of
% building the block collection $B$ is insignificant, as it typically requires a single
% iteration over the input profiles, $O(|P|)$. Finally, the time complexity of its
% emission phase is usually constant, unless its Comparison List gets empty. In
% these cases, \textsf{PBS} refills its Comparison List with the sorted
% comparisons of the next block to processed. The time complexity of this
% procedure is very low, 
% %$O(\|\bar{b}\|$$\cdot$$\log\|\bar{b}\|)$ on average,
% since it is dominated by the sorting of all comparisons in an individual block,
% i.e., $O(\|\bar{b}\|$$\cdot$$\log\|\bar{b}\|)$ on average, rather than the
% sorting of the entire block collection, $B$.

%{\color{blue}
\subsubsection{Progressive Profile Scheduling (\textbf{\textsf{PPS}})}
\label{sec:pps}
The block-centric functionality of \textsf{PBS} is crafted for an Edge Weighting approach that operates at the level of individual comparisons.
We now propose a novel progressive method with entity-centric
functionality, called Progressive Profile Scheduling (\textbf{\textsf{PPS}}).

\textsf{PPS} is based on the 
concept of \textit{duplication likelihood}, i.e., the likelihood of an
individual profile $p_i$ to have matches.
In \textsf{Clean-clean ER}, the duplication likelihood of $p_i\in P_1$ corresponds
to its likelihood to have a match in $P_2$, since there can be up to one
matching profile per entity in every profile collection.
In \textsf{Dirty ER}, though, the duplication likelihood of $p_i$ is
analogous to the size of its equivalence cluster, i.e., high values indicate
that $p_i$ matches with many other profiles, and vice versa for low values.

In fact, \textsf{PPS} aims to sort all profiles in decreasing
duplication likelihood, forming a data structure that is called
\textbf{Sorted Profile List}. Then, moving from the top to the
bottom of this list, \textsf{PPS} goes iteratively through every profile,
emitting the top-$k$ weighted comparisons that entail it in decreasing matching likelihood.

To build the Sorted Profile List, \textsf{PPS} derives the duplication
likelihood of every profile from a given Blocking Graph. 
The underlying assumption is the same as for all methods based on a Blocking Graph: the weight of a blocking
graph edge captures the matching likelihood between the
adjacent profiles. Thus, the duplication likelihood of each node (i.e., profile)
is estimated by aggregating the weights of its incident edges. In particular,
our implementation of \textsf{PPS} approximates the duplication likelihood of a
profile through the average weight of the edges that are incident to the
corresponding node - other aggregation functions can be
employed instead, but the \textit{average} one consistently exhibited high
performance across different datasets.

\begin{algorithm}[t]
{\scriptsize
  \LinesNumbered %
  \SetAlgoLined %SetLine
  \SetAlgoVlined %SetVline 
  \KwIn{(i)~Profile~collection:~$P$,~(ii)~Weighting~scheme,~$wScheme$}
  \KwOut{The overall best comparison}
  $B$ $\leftarrow$ buildRedundancyPositiveBlocks($P$)\;
  $ProfileIndex$ $\leftarrow$ buildProfileIndex($B$)\;
  $SortedProfileList$ $\leftarrow$ $\emptyset$\;
  $topComparisonsSet$ $\leftarrow$ $\emptyset$\;
  \ForEach{$p_i$ $\in$ $P$}{
    $weights$[] $\leftarrow$ $\emptyset$\;
  	$distinctNeighbors$ $\leftarrow$ $\emptyset$\;
  	\ForEach{$b_k$ $\in$ $ProfileIndex$.getBlocks($p_i$)}{
  		\ForEach{$p_j$($\neq$$p_i$) $\in$ $b_k$}{
  			$weights$[$j$]~+=~wScheme($p_j$, $p_i$, $b_k$)\;	
  			$distinctNeighbors$.add($j$)\;
  		}
  	}
  	
  	$topComparison$ $\leftarrow$ $null$\;
  	$duplicationLikelihood$ $\leftarrow$ 0\;
  	\ForEach{$j$ $\in$ $distinctNeighbors$} {
  		$duplicationLikelihood$~+=~$weights$[$j$]\;
  		\If {$topComparison$.getWeight() $<$ $weights$[$j$]} { 
  			$topComparison$ $\leftarrow$ getComparison($i$, $j$, $weights$[$j$]\;
  		}
	}
	$topComparisonsSet$.add($topComparison$)\;
	
	$duplicationLikelihood$~/=~$distinctNeighbors$.size()\;
	$SortedProfileList$.add($p_i$, $duplicationLikelihood$)\;
  }
	
  $ComparisonList$.addAll($topComparisonsSet$)\;
  sortInDescreasingWeight($ComparisonList$)\;
  sortInDescreasingWeight($SortedProfileList$)\;
	
  \KwRet{$ComparisonList$.removeFirst()}\;
  \caption{{\small Initialization phase for \textsf{PPS}.}}
  \label{alg:initCEP}
}
\end{algorithm}
% \begin{algorithm}[t]
% {\scriptsize
%   \LinesNumbered %
%   \SetAlgoLined %SetLine
%   \SetAlgoVlined %SetVline
% %   \KwIn{The weight of the last comparison, $w_{last}$} 
%   \KwOut{The next best comparison}
%   \If {$ComparisonList$.isEmpty()} {
%   	\tcc{repeat lines 3 - 15 in Algorithm \ref{alg:initCEP}}
%   	\ForEach{$j$ $\in$ $distinctNeighbors$} {
%   		$c_{i,j}$ $\leftarrow$ getComparison($i$, $j$, $weights$[$j$])\;
%   		\If {$c_{i,j}$.getWeight() $<$ $w_{last}$} {
%   			$SortedStack$.push($c_{i,j}$)\;
%   		} 
%   		\If {$K_{max}$ $<$ $SortedStack$.size()} {
%   			$SortedStack$.pop()\;
%   		}
% 	}
% 	$ComparisonList$	$\leftarrow$ sortDescreasingWeight($SortedStack$)\;
%   }
%   $c_{first}$ $\leftarrow$ $ComparisonList$.removeFirst()\;
%   $w_{last}$ $\leftarrow$ $c_{first}$.getWeight()\;
%   \KwRet{$c_{first}$}\;
%   \caption{{\small Emission phase for \textsf{R-CEP}.}}
%   \label{alg:emitCEP}
% }
% \end{algorithm}
\begin{algorithm}[t]
{\scriptsize
  \LinesNumbered %
  \SetAlgoLined %SetLine
  \SetAlgoVlined %SetVline
%   \KwIn{The weight of the last comparison, $w_{last}$} 
  \KwOut{The next best comparison}
  $checkedEntities$ $\leftarrow$ $\emptyset$\;
  \If {$ComparisonList$.isEmpty()} {
  \If {$SortedProfileList$.isNotEmpty()} {
  	$p_i$ = $SortedProfileList.removeFirst()$\;
  	$checkedEntities$.add($i$)\;
	$weights$[] $\leftarrow$ $\emptyset$\;
  	$distinctNeighbors$ $\leftarrow$ $\emptyset$\;
	$SortedStack$ $\leftarrow$ $\emptyset$\;
  	\ForEach{$b_k$ $\in$ $ProfileIndex$.getBlocks($p_i$)}{
  		\ForEach{$p_j$($\neq$$p_i$) $\in$ $b_k$}{
  			\If {$checkedEntities$.contains($j$)} {
  				continue\;
  			}
  			$weights$[$j$]~+=~wScheme($p_j$, $p_i$, $b_k$)\;	
  			$distinctNeighbors$.add($j$)\;
  		}
  	}
  	
  	\ForEach{$j$ $\in$ $distinctNeighbors$} {
  		$SortedStack$.push(getComparison($i$, $j$, $weights$[$j$])\;
  		\If {$K_{max}$ $<$ $SortedStack$.size()} {
  			$SortedStack$.pop()\;
  		}
	}
% 	$SortedStack$.pop()\;
	$ComparisonList$	$\leftarrow$ sortInDescreasingWeight($SortedStack$)\;
   }
  }
  \KwRet{$ComparisonList$.removeFirst()}\;
  \caption{{\small Emission phase for \textsf{PPS}.}}
  \label{alg:emitCEP-pList}
}
%\vspace{-20pt}
\end{algorithm}

During the creation of the Sorted Profile List, \textsf{PPS} also initializes
the Comparison List with the set of the top-weighted comparisons of each node.
This step does not require any additional computational cost. While investigating the
neighborhood of a particular node, \textsf{PPS} retains in a local
variable the highest edge weight along with the corresponding comparison.
After traversing all edges in the neighborhood, the overall best comparison is
added to the set $topComparisonsSet$. As soon as all nodes have been processed,
\textsf{PPS} sorts the elements of $topComparisonsSet$ in decreasing matching
likelihood and adds them to the Comparison List. In the end, this process
allows for emitting the comparison with the highest weight across the entire
Blocking Graph, i.e., the comparison placed in the first position of the
Comparison List.

In more detail, the initialization phase of \textsf{PPS} is outlined in
Algorithm \ref{alg:initCEP}. First, a redundancy block collection is created along with the
corresponding $ProfileIndex$ (Lines 1-2). Subsequently, \textsf{PPS} iterates
over all input profiles (Line 5) and for every profile $p_i$, it goes through
the blocks that contain it, $B_i$, which are derived from the Profile Index
(Line 8). For every such block, it iterates over the co-occurring profiles, placing
them into the set of neighbor ids and updating their overall weight
(Lines 9-11). After examining all blocks, it goes through the set of neighbor
profiles in order to estimate the overall duplication likelihood and identify
the top-weighted comparison (Lines 12-17). The selected comparison is then added to
the set of top-weighted comparisons, which makes sure that none of them
is repeated, while the current profile is added to the Sorted Profile List along
with its duplication likelihood (Lines 18-20). After processing all profiles,
the top-weighted comparisons are added to the Comparison List to be sorted
in decreasing order of weights; the same applies to Sorted Profile List (Lines
21-23). Finally, the overall top-weighted comparison is emitted (Line 24).

The emission phase of \textsf{PPS} relies on two pillars:

\emph{(i)} A data structure called $SortedStack$, which
contains a set of comparisons such that they are constantly sorted in
non-decreasing weight, from the lowest to the highest one. Thus, its head always
corresponds to the comparison with the lowest weight and can be efficiently
removed with the a \textit{pop} operation of constant computational cost,
$O(1)$.

\emph{(ii)} A custom mechanism for avoiding repeated comparisons that relies
on a set with all entities that have already been processed, called 
$checkedEntities$. Before considering the comparison of the current profile $p_i$
with a co-occurring one $p_j$, $c_{ij}$, we investigate whether
$checkedEntities$ contains the id $j$. If yes, $c_{ij}$ is skipped, based on the observation that the
most important comparisons of $p_j$ have already been emitted. In this way, we
disregard even comparisons that are among the $K_{max}$ top-weighted ones for
the current entity, $p_i$, but not for the previously examined one, $p_j$.
The reason is that $p_j$'s higher duplication likelihood provides more
reliable evidence for  $c_{ij}$'s low matching likelihood. 

\begin{figure}[t!]\centering
%\color{blue}
	\includegraphics[width=0.45\textwidth]{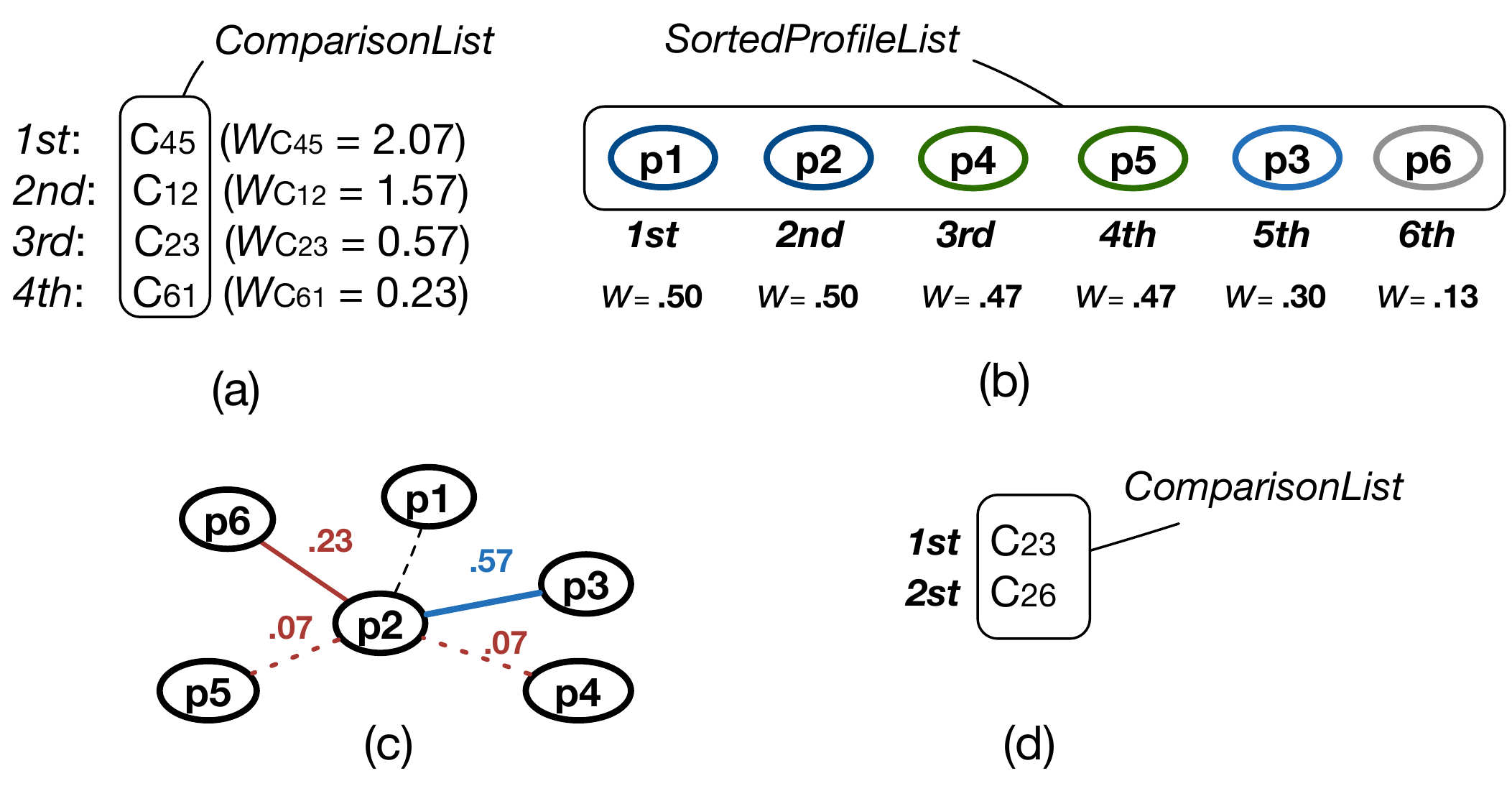}
	\vspace{-12pt}
	\caption{(a) The Comparison List after the initialization phase, containing the
	top-weighted comparison of every node in the Blocking Graph of Figure
	\ref{fig:example}c. (b) The corresponding Sorted Profile List.
	(c) The node neighborhood of $p_{2}$ in the same Blocking Graph.
	(d) The Comparison List after processing $p_{2}$ during the emission phase.}
	\vspace{-12pt}
	\label{fig:pps}
\end{figure}

\begin{table*}[t]
\centering
%\scriptsize
\small
\caption{Space and time complexities of our methods.
%$\bar{s_e}$ is the average number of suffixes per profile.
In some cases, the emission phase complexity 
% phase correspond to different may
% have different time complexities, 
depends on the status of the Comparison List.}
\setlength{\tabcolsep}{2.1pt}
\vspace{-8pt}
\renewcommand{\arraystretch}{1.3}
\scriptsize
\begin{tabular}{|l|l|c|c|c|}
		   \hline
           \multicolumn{1}{|c|}{\multirow{2}{*}{Method}} &
           \multicolumn{1}{c|}{\multirow{2}{*}{Acronym}} & 
           Space & \multicolumn{2}{c|}{Time complexity}\\
           \cline{4-5} 
           & & Complexity & Initialization Phase & Emission
           Phase\\
           \hline
           \hline
		   Schema-agnostic Progressive Sorted Neighborhood &
		   \textsf{SA-PSN} & 
		   $O($$\kpp$$\cdot$$|P|)$ &
		   $O($$\kpp$$\cdot$$|P|$$\cdot$$\log$$(\kpp$$\cdot$$|P|))$ & $O(1)$
		   \\
		   \hline
		   Schema-agnostic Progressive Suffix Arrays Blocking &
		   \textsf{SA-PSAB} &
		   $O($$\bar{s_e}$$\cdot$$|P|)$ &
		   $O($$\bar{s_e}$$\cdot$$|P|$$\cdot$$\log$$(\bar{s_e}$$\cdot$$|P|))$ & $O(1)$
		   \\
		   \hline
		   \hline
		   Global Scheam-agnostic Progressive Sorted Neighborhood &
		   \textsf{GS-PSN} &  
		   $O(w_{max}$$\cdot$$\kpp$$\cdot$$|P|)$ & 
		   $O($$\kpp$$\cdot$$|P|$$\cdot$$\log$$(\kpp$$\cdot$$|P|))$ & $O(1)$
		   \\
		   \hline
		   Local Scheam-agnostic Progressive Sorted Neighborhood &
		   \textsf{LS-PSN} & 
		   $O(\kpp$$\cdot$$|P|)$ &
		   $O($$\kpp$$\cdot$$|P|$$\cdot$$\log$$(\kpp$$\cdot$$|P|$))
		   & $O(1)$ or $O($$\kpp$$\cdot$$|P|)$ \\
		   \hline
		   Progressive Profile Scheduling &
		   \textsf{PPS} &  
		   $O(\kpp$$\cdot$$|P|)$ & $O(|V_B|+|E_B|)$ &
		   $O(1)$ or $O(\kpp$$\cdot$$\ppb)$\\
 		   \hline
		   Progressive Block Scheduling &
		   \textsf{PBS} &  
		   $O(\kpp$$\cdot$$|P|)$ &
		   $O(|B|$$\cdot$$\log|B|)$ & 
		   $O(1)$ or $O(\|\bar{b}\|$$\cdot$$\log\|\bar{b}\|)$ \\
		   \hline
\end{tabular}
%\vspace{-10pt}
%When two different time complexities of the emission phase are reported, both are valid under different circumstances.}
\label{table:complexities}
\vspace{-12pt}
\end{table*}

In more detail, the emission phase of \textsf{PPS} is outlined in Algorithm
\ref{alg:emitCEP-pList}. Initially, it emits the top-weighted comparisons that
were placed in the Comparison List during initialization. As soon as this list gets empty, \textsf{PPS}  
iterates over the individual profiles according to their duplication
likelihood, from the highest to the lowest one (Lines 2-4). For the next
available profile, \textsf{PPS} retrieves the associated blocks from the Profile
Index (Line 9) and iterates over their contents in order to gather their 
$K_{max}$ top-weighted comparisons (Lines 10-19): initially, \textsf{PPS} goes
through the co-occurring profiles, skipping the already examined ones (Lines 10-12). The
non-examined ones are then added to the set of neighbor ids and their overall
weight is updated (Lines 13-14). After examining all blocks, \textsf{PPS}
estimates the overall weight for every neighbor, pushing the corresponding 
comparison in the sorted stack (Lines 15-16). If the size of the stack 
exceeds $K_{max}$, the comparison with the lowest weight is popped (Lines
17-18). 

Finally, the remaining comparisons are sorted in decreasing weights and placed in the Comparison List (Line 19), followed by emission of the top-weighted comparison (Line 20).

\begin{ex}
To illustrate the functionality of \textsf{PPS}, consider the example in Figure
\ref{fig:pps}.
During the initialization phase, \textsf{PPS} iterates over all nodes of the
Blocking Graph to compute the average weight of the incident edges along with
the top-weighted comparison in every node neighborhood.
At the end of this iteration, all top-weighted comparisons and all
profiles are sorted in non-increasing weights, from the highest to the lowest
one, in order to form the Comparison List in Figure \ref{fig:pps}a and the
Sorted Profile List in Figure \ref{fig:pps}b, respectively.
During the emission phase, \textsf{PPS} initially emits all comparisons in the
Comparison List of Figure \ref{fig:pps}a. Then, it goes through the Sorted
Profile List, one node at a time, gathering the top-k comparisons in the corresponding node neighborhood.
For instance, Figure \ref{fig:pps}c shows the neighborhood of $p_{2}$, whose top-2 edges are inserted in the Comparison List of Figure \ref{fig:pps}d.
Note that $p_{1}$ has already been processed, since it was placed first in the
Sorted Profile List of Figure \ref{fig:pps}b.
As a result, the control in Line 11 in Algorithm \ref{alg:emitCEP-pList},
checkedEntities.contains(1), returns \texttt{true} and $c_{12}$ is not inserted
in the Comparison List of Figure \ref{fig:pps}d, despite its high edge weight.
%since $p_1$ has already been emitted, i.e.,

\end{ex}

% \textbf{Complexity Analysis.} The space requirements of \textsf{PPS} are
% dominated by the size of the Profile Index, which associates every profile with
% the corresponding block ids. Therefore, they are equal to
% $O($$\kpp$$\cdot$$|P|)$, on average, depending linearly on the size of the
% input. Its initialization time complexity is $O(|V_B|+|E_B|)$, as \textsf{PPS}
% iterates over all nodes and edges of the blocking graph $G_B$, without any pruning. For the
% emission phase, the time complexity is constant in most cases, simply emitting
% the next best comparison. Whenever the Comparison List gets empty, though,
% \textsf{PPS} sorts the comparisons associated with the next best entity in
% non-increasing matching likelihood, with a computational cost that amounts to
% $O($$\kpp$$\cdot$$|P|)$, on average.

\vspace{-10pt}
\section{Complexity Analysis}
\label{sec:complexities}

We now elaborate on the space and time complexities of all
algorithms presented in Sections \ref{sec:naiveSolutions} and
\ref{sec:approach}. All complexities are summarized in Table
\ref{table:complexities}.

\vspace{-9pt}
\subsection{Space Complexity.}
\vspace{-2pt}
We observe that for most methods, the space complexity is linear with respect to
the size of the input dataset, $|P|$. 
For \textsf{SA-PSN} and \textsf{LS-PSN},
it is just $O($$\kpp$$\cdot$$|P|)$, where $\kpp$ is the average
number of name-value pairs ($\sim$ blocking keys per entity),
because they mainly keep in memory the Profile List. \textsf{LS-PSN} additionally maintains the Position Index, but it
has exactly the same complexity. The same holds for the Profile Index, which 
dominates the space requirements of \textsf{PPS} and \textsf{PBS}.
\textsf{GS-PSN} occupies more space, $O(w_{max}{\cdot}\kpp{\cdot}|P|)$, due
to the Comparison List, which, in the worst case, contains 1 comparison per
position in the Profile List for every window size. To keep the suffix forest in
memory, \textsf{SA-PSAB} has a space complexity of $O($$\bar{s_e}$$\cdot$$|P|)$,
where $\bar{s_e}$ is the average number of suffixes per profile.
%Finally, the space requirements of \textsf{R-CEP} and \textsf{NC-CEP} are practically determined by the available memory resources.
% Finally, the lowest complexity corresponds to \textsf{CEP-pList}, as it simply keeps a
% sorted list of profiles in memory. 
Thus, we can conclude that all methods scale well to the Volume of Web data.

\vspace{-10pt}
\subsection{Time Complexity}
\vspace{-2pt}
\textbf{Initialization phase.}
All methods are also scalable 
% Equally scalable are all methods 
with respect to the time complexity of their
initialization phase. For the similarity-based methods,
\textsf{SA-PSN}, \textsf{LS-PSN} and \textsf{GS-PSN}, the time complexity is
dominated by the sorting of blocking keys in alphabetical order, 
$O($$\kpp$$\cdot$$|P|$$\cdot$$\log$$(\kpp$$\cdot$$|P|))$. 
For \textsf{SA-PSAB}, it is
$O($$\bar{s_e}$$\cdot$$|P|$$\cdot$$\log$$(\bar{s_e}$$\cdot$$|P|))$, as it sorts
all suffixes (i.e., tree nodes) in non-increasing order of length and
non-decreasing order of comparisons.
For \textsf{PPS}, the initialization time complexity is $O(|V_B|+|E_B|)$,
as this method iterates over all nodes and edges of the blocking graph $G_B$,
without any pruning. 
% comparisons of the redundancy-positive block
% collection $B$ in order to estimate all edge weights.
% Finally, for \textsf{PBS} applies edge weighting locally, to the comparisons of
% individual blocks.
Finally, the time complexity of \textsf{PBS} is dominated by the sorting of
blocks in non-decreasing comparisons, i.e., $O(|B|{\cdot}\log|B|)$.
Note that for both equality-based methods, the cost of building the block
collection $B$ is insignificant, as it typically requires a single iteration
over the input profiles, $O(|P|)$.

\textbf{Emission  phase.}
%  the time complexity of the emission phase, 
We distinguish all methods into three categories with respect to the time
complexity of this phase. The first one includes the na\"ive methods,
\textsf{SA-PSN} and \textsf{SA-PSAB}, which simply return the next comparison in a window or tree
node, thus exhibiting a constant time complexity, $O(1)$. Yet, a large part
of the emitted comparisons is repeated, as every profile is associated with
multiple keys. Due to their simplicity, though, these methods make no provision
for detecting repeated comparisons. 

The second category includes
\textsf{GS-PSN}, which exhibits a constant time complexity, $O(1)$, without
emitting repeated comparisons. The reason is that it precomputes all
comparisons, discarding the repeated ones.

The third category involves all methods with unstable response time, namely
\textsf{LS-PSN}, \textsf{PPS} and \textsf{PBS}.
In most cases, their time complexity is constant, but whenever their Comparison
List gets empty, they renew its contents by repeating (part of) their
initialization phase. In fact, the emission time complexity of
\textsf{LS-PSN} is equal to that of the initialization phase, as the same
process is applied to the entire Sorted Profile List; the only difference is
the incremented window size. \textsf{PBS} also applies the same procedure as
the initialization phase in order to refill its Comparison List. Yet, the time
complexity is now much lower, as it is dominated by the sorting of all
comparisons in an individual block, i.e., $O(\|\bar{b}\|$$\cdot$$\log\|\bar{b}\|)$, on
average, rather than by the sorting of the entire block collection, $B$.
Finally, the emission phase of \textsf{PPS} is significantly more efficient than
its initialization phase: it merely sorts the comparisons associated with a single entity in
non-increasing matching likelihood, $O($$\kpp$$\cdot$$|P|)$, on
average, instead of sorting all profiles in non-increasing duplication
likelihood.

\section{Experiments}
\label{sec:experiments}

\begin{figure*}[ht!]
	\centering
	\includegraphics[width=1\textwidth]{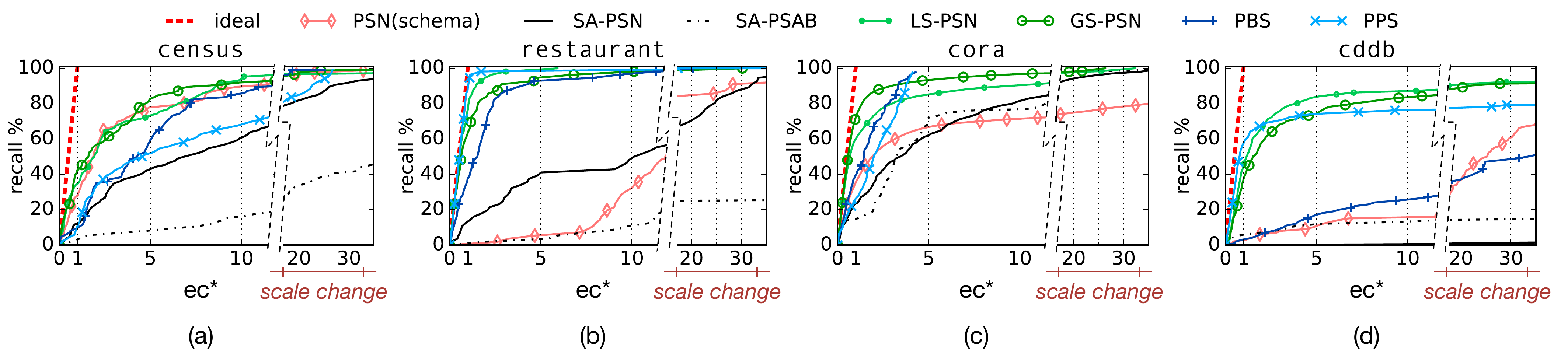}
	\vspace{-24pt}
	\caption{Recall progressiveness over the
	\textsf{structured} datasets.}
	\vspace{-14pt}
	\label{fig:small}
\end{figure*} 

\begin{figure}[t!]\centering
	\includegraphics[width=0.45\textwidth]{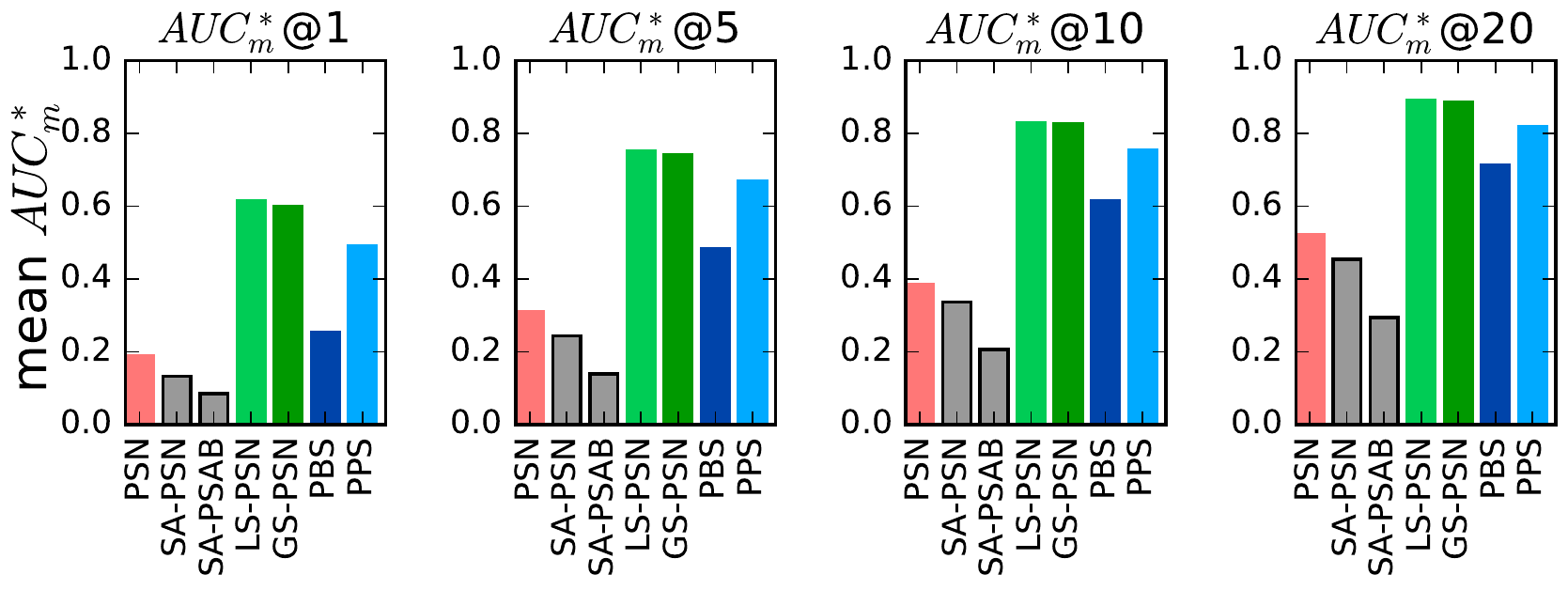}
	\vspace{-12pt}
	\caption{Mean $AUC_m^*$ over the \textsf{structured} datasets.}
	\vspace{-15pt}
	\label{fig:auc_small}
\end{figure}

\noindent
\textbf{Datasets.}
For the experimental evaluation, we employ 7 diverse real-world datasets that are widely
adopted in the literature as benchmark data for \textsf{ER}
\cite{DBLP:journals/tkde/Christen12, DBLP:conf/edbt/0001PPK16,
DBLP:journals/pvldb/SimoniniBJ16, DBLP:journals/pvldb/KopckeTR10, btc-2009}.
Their characteristics are reported in Table \ref{table:datasets}.
The \texttt{census}, \texttt{restaurant}, \texttt{cora}, and \texttt{cddb} datasets are extracted from a single data source
containing duplicated profiles, hence they are meant to test \textsf{Dirty ER}
tasks.
The remaining datasets (\texttt{movies}, \texttt{dbpedia}, and \texttt{freebase}) are suitable for testing scalability, as well as \textsf{Clean-clean ER}, since they are extracted from two different data sources, where matching profiles exist only between a source and another:
\texttt{movies} from \texttt{imdb.com} and \texttt{dbpedia.org};
\texttt{dbpedia} from two different snapshots of DBpedia (\texttt{dbpedia.org}
2007-2009)\footnote{Due to the constant changes in DBpedia, the two versions
share only 25\% of the \textit{name-value} pairs, forming an non-trivial
\textsf{ER} task \cite{DBLP:journals/pvldb/0001APK15,
DBLP:conf/edbt/0001PPK16}.}; \texttt{freebase} from
\texttt{developers\-.google.com/freebase/} and \texttt{dbpedia.org} (extracted from \cite{btc-2009}).
For all the datasets, the ground truth is known and provided with the data.

\begin{table}[]
\centering
\scriptsize
\setlength{\tabcolsep}{3pt}
%\small
\caption{Dataset characteristics: \textsf{ER} type, number of entity profiles
($|P|$), number of \textit{attribute names}, number of existing matches
($|\mathcal{D}_P|$) and average number of name-value pairs per entity
($|\bar{p}|$).
%{\bf ??? XXXXXXXXXXXXX table goes out of bounds! XXXXXXXXXXXXX ???}
}
\vspace{-7pt}
\begin{tabular}{|c|c|c|c|c|c|}
\cline{2-6}
\multicolumn{1}{c|}{} & \textsf{ER} type     & $|P|$      & \#attr.     &
$|\mathcal{D}_P|$ & $|\bar{p}|$\\
\hline
\multicolumn{6}{c}{Structured Datasets}\\
\hline

\texttt{census}     & \textsf{Dirty ER}       & 841       & 5  & 344 & 4.65 \\
\texttt{restaurant} & \textsf{Dirty ER}       & 864       & 5 & 112 & 5.00 \\
\texttt{cora}       & \textsf{Dirty ER}       & 1.3k      & 12  & 17k & 5.53\\
\texttt{cddb}       & \textsf{Dirty ER}       & 9.8k      & 106 & 300 & 18.75\\

\hline
\multicolumn{6}{c}{Large, Heterogeneous Datasets}\\
\hline
%\texttt{articles}   & \textsf{Clean-clean ER} & 2.6k-2.3k & 4-4 & 2.2k   \\
%\texttt{products}   & \textsf{Clean-clean ER} & 1.1k-1.1k & 4-4 & 1.1k   \\
\texttt{movies}     & \textsf{Clean-clean ER} & 28k---23k   & 4---7  & 23k &
7.11\\
\texttt{dbpedia}    & \textsf{Clean-clean ER} & 1.2M---2.2M & 30k---50k & 893k
& 15.47\\
\texttt{freebase}    & \textsf{Clean-clean ER} & 4.2M---3.7M & 37k---11k & 1.5M
& 24.54\\

\hline

\end{tabular}
\vspace{5pt}
\label{table:datasets}
\vspace{-13pt}
\end{table}
%\vspace{2pt}

%In each \textit{structured} dataset, profiles abide by a known schema;
%Hence allowing the definition of a schema-based blocking key the definition of schema-based blocking criterionand the \textit{best blocking key} for \textsf{PSN} are known from the literature \cite{DBLP:journals/tkde/Christen12}.

For the \textit{structured} datasets, the \textit{best schema-based blocking
keys} for \textsf{PSN} are known from the
literature~\cite{DBLP:journals/pvldb/0001APK15,DBLP:journals/tkde/Christen12}\footnote{See
also the code at:
\href{https://sourceforge.net/projects/febrl}{https://sourceforge.net/projects/febrl}
and
\href{https://sourceforge.net/projects/erframework}{https://sourceforge.net/projects/erframework}.}.
Note that the schema-based methods are inapplicable to the \textit{large, he\-te\-ro\-ge\-neous} datasets.
This is due to the size of the attribute set and the lack of a schema-alignment
for \textsf{Clean-clean} datasets.
%\footnote{\textsf{movies} has a total of 11 distinct
%attributes, but to the best of our knowledge no schema-based blocking key is
%known from the literature to perform well, while the schema-alignment for
%determining a schema-based blocking key is non-trivial.}
Finally, in \textsf{dbpedia} and \textsf{freebase}, there is a very small
overlap in the attributes describing their profile collections.

\vspace{2pt}
\noindent
\textbf{System setup.}
All methods are implemented in Java 8 and the code is publicly
available\footnote{\url{https://stravanni.github.io/progressiveER/}}.
All experiments have been performed on a server running Ubuntu 14.04, with 80GB RAM, and an Intel
Xeon E5-2670 v2 @ 2.50GHz CPU.
Note that we limited the maximum heap size parameter of the JVM to
8GB for the structured datasets and for \texttt{movies}, while for
\texttt{DBPedia} and \texttt{Freebase} we set that parameter to 80GB.

\vspace{2pt}
\noindent \textbf{Parameter configuration.}
We apply the following settings to all datasets.
%for \textsf{SA-PSAB}, we set $l_{min}$=2.
For \textsf{GS-PSN}, we set $w_{max}$=20 for structured datasets and $w_{max}$=200 for large, heterogeneous datasets|preliminary experiments have shown that these values work for all
the datasets. For \textsf{PBS} and \textsf{PPS}, we can use any schema-agnostic
blocking method that produces redundancy-positive blocks, like DisNGram
\cite{DBLP:journals/tkde/SongLH17}.

We opted for the \textit{Token
Blocking Workflow}, which
%  to derive the redundancy-positive block collection.
% This workflow 
has been experimentally verified to address effectively and
efficiently the Volume and Variety of Web data \cite{DBLP:conf/edbt/0001PPK16}.
It consists of the following steps: \emph{(1)} Schema-agnostic Standard Blocking
\cite{DBLP:journals/pvldb/0001APK15} 
%a.k.a. Token Blocking, 
creates a separate
block for every attribute value token that stems from at least two profiles.
\emph{(2)} \textit{Block Purging} \cite{DBLP:conf/edbt/0001PPK16} discards large
blocks that correspond to stop words, involving more than 10\% of the input
profiles.
\emph{(3)} \textit{Block Filtering} \cite{DBLP:conf/edbt/0001PPK16} retains
every profile in 80\% of its most important (i.e., smallest) blocks. \emph{(4)}
\textsf{ARCS}  performs edge weighting on the Blocking Graph.

\vspace{2pt}
\noindent
\textbf{Metrics.}
\textsf{Recall} is typically employed to evaluate the effectiveness of a \textsf{Batch ER} method $m$ over a profile collection $P$.
It measures the portion of
detected matches: $recall {=} |\mathcal{D}_m|/|\mathcal{D}_P|$,
where $\mathcal{D}_m$ is the set of matches detected (emitted) by $m$,
while $\mathcal{D}_P$ is the set of all matches in $P$.

In \textsf{Progressive ER}, we are interested in how fast matches are
emitted. To illustrate this, we consider \textbf{recall progressiveness} by
plotting the evolution of recall (vertical axis) with respect to the
\textit{normalized number of emitted comparisons} (horizontal axis): $ec^*{=}ec/|\mathcal{D}_P|$,
where $ec$ is the number of emitted comparisons at a certain time during the
processing. The purpose of this normalization is twofold: \emph{(i)} it allows
for using the same scale among different datasets, and \emph{(ii)} it
facilitates the comparison of all progressive methods with the \textit{ideal}
one, which achieves recall=1 after emitting just the first $|\mathcal{D}_P|$
comparisons, i.e., at $ec^*{=}1$.

To facilitate the comparisons between progressive methods, we quantify their
progressive recall using the \textbf{area under the curve} (AUC) of the
above plot (the AUC expressed in function of $ec$ - not the normalized
$ec^*$ - is known in the literature as \textit{progressive recall}
\cite{DBLP:journals/pvldb/FirmaniSS16}, and is employed for the same purpose).
For a method $m$, we indicate with $AUC_m@ec^*$ the value of AUC for a given
$ec^*$; for instance, $AUC_{PSN}@5$ is the area under the recall curve of the
method \textsf{PSN} after the emission of $ec{=}5{\cdot}|\mathcal{D}_P|$ comparisons.
To restrict $AUC_m@ec^*$ to the interval $[0,1]$, we normalize it with the
performance of the ideal method: $AUC_m^*@ec^* =
\frac{AUC_m@ec^*}{AUC_{ideal}@ec^*}$. 
$AUC_m^*@ec^*$ is called \textbf{normalized area under the curve}:
higher values correspond to a better \textit{progressiveness}, with the ideal
method having $AUC_{ideal}^*{=}1$ for any value of $ec^*$.

For the time performance evaluation of a method $m$, we consider the
\textbf{initialization time} and the \textbf{comparison time}:
the former is the time required to emit the first comparison, considering all
the pre-processing steps (e.g., Schema-agnostic Standard Blocking, Block Purging, Block Filtering
for \textsf{PBS}); the comparison time is the average time between two
consecutive comparison emissions. It includes both the emission time (i.e., the
time required for generating the next best comparison) and the time required for
applying the selected match function to that comparison.

\vspace{2pt}
\noindent \textbf{Baselines.}
In the following, we use \textsf{PSN} and \textsf{SA-PSN} as baseline methods.
As explained above, the best schema-based blocking keys, which are
necessary for \textsf{PSN}, are only known for the \textsf{Dirty ER} datasets. 
%for \textsf{PSN} are known from \cite{DBLP:journals/tkde/Christen12}.
For the \textsf{Clean-clean ER} ones, no such blocking keys have been reported
in the literature. 
% blocking keys.
As a result, we consider only \textsf{SA-PSN} as baseline method for
\textsf{Clean-clean ER} datasets.
%Finally, for \textsf{PPS}, we set $K_{max}{=}10{\cdot}\kpp$, since preliminary experiments verified that it yields good performance\footnote{Yet, similar heuristics (e.g., $K_{max}{=}5{\cdot}\kpp$) yield very similar performances.}.

\vspace{-10pt}
%\subsection{Schema-based vs. Schema-agnostic Progressive ER}
\subsection{Structured Datasets}
\label{sec:small}
\vspace{-2pt}
\begin{figure*}[ht!]
	\centering
	\includegraphics[width=0.97\textwidth]{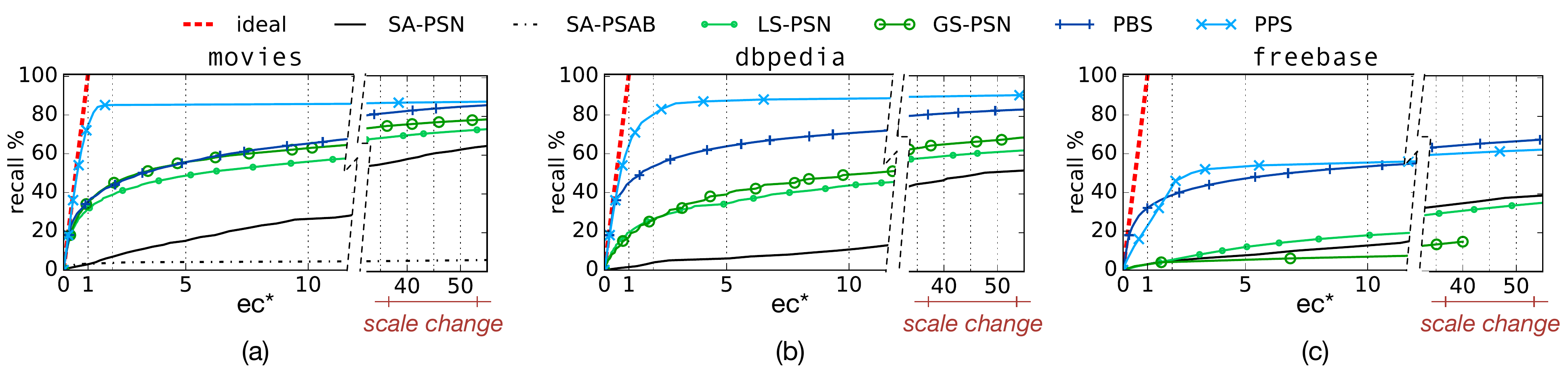}
	%\includegraphics[width=0.8\textwidth]{images/exp/large_v_20}
	%\vspace{-13pt}
	\vspace{-14pt}
	\caption{Recall progressiveness over the \textsf{large}, \textsf{heterogeneous}
	datasets.}
	\vspace{-18pt}
	\label{fig:large}
\end{figure*}

\begin{figure}[t!]\centering
	\includegraphics[width=0.47\textwidth]{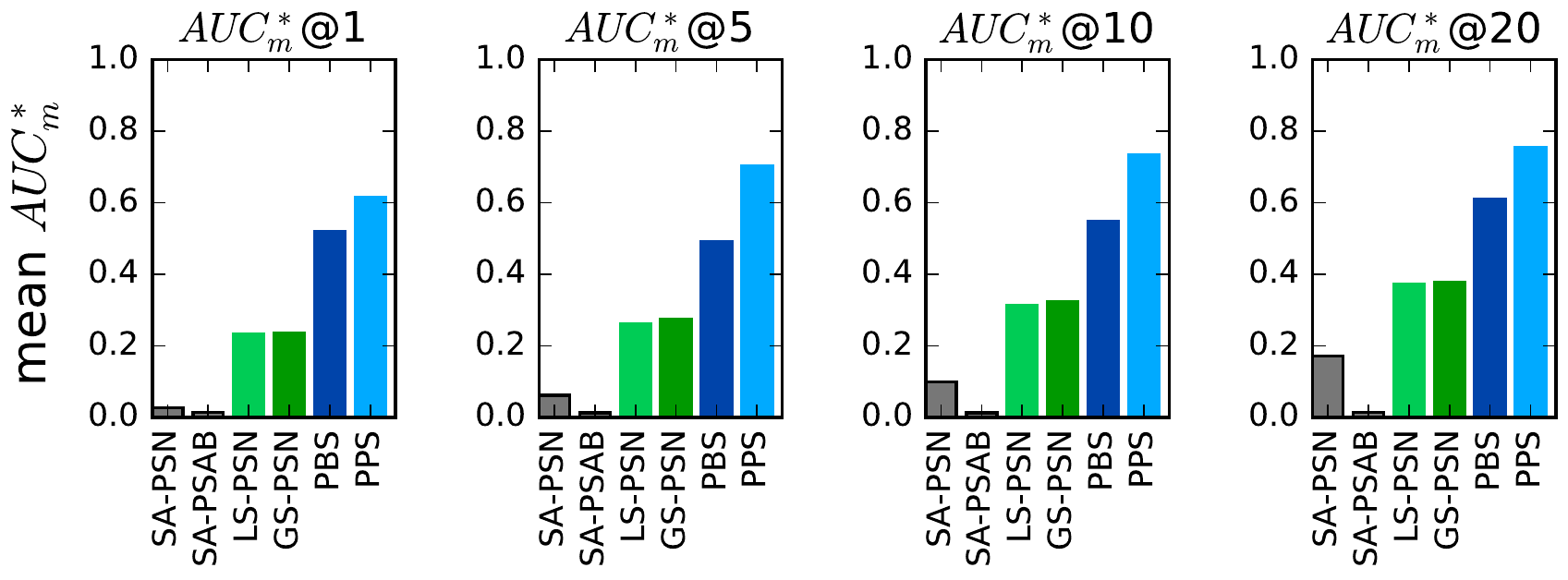}
	\vspace{-13pt}
	\caption{Mean  $AUC_m^*$ over the \textsf{large}, \textsf{heterogeneous} datasets.}
	\vspace{-8pt}
	\label{fig:auc_large}
\end{figure}

We now compare our schema-agnostic methods against the state-of-the-art
schema-based method, i.e. \textsf{PSN}
\cite{DBLP:journals/tkde/PapenbrockHN15,DBLP:journals/tkde/WhangMG13}, on the
\textsf{structured} datasets.
% {\color{red}--- please remember that for the \textsf{large}, \textsf{heterogenous} ones PSN is inapplicable. IS THIS SENTENCE USEFULL??}
We assess the relative effectiveness of all methods with respect to
recall progressiveness.
The corresponding plots appear in Figure \ref{fig:small}.
They depict the performance of all methods for up to $ec^*{=}30$, i.e., we
measure the recall for a number of comparisons thirty times the comparisons
required by the ideal method to complete each \textsf{ER} task. We focus,
though, on the interval [0,10]
% ,  Figure \ref{fig:small} focuses on the 0 to 10 range, 
so as to highlight the behavior of the methods in the early stage of
\textsf{ER}, the most critical for pay-as-you-go applications.

We observe that the advanced schema-agnostic methods outperform \textsf{PSN}, \textsf{SA-PSN} and \textsf{SA-PSAB} across all data\-sets\footnote{In Figure \ref{fig:small}d, the curve of \textsf{SA-PSN} is too low to be visible, almost coinciding with the horizontal axis.}.
Only for \texttt{census} does \textsf{PSN} perform better than \textsf{PBS}
% The only dataset on which \textsf{PSN} performs better than \textsf{PBS} 
(but not better that \textsf{LS/GS-PSN})  -- see
Figure~\ref{fig:small}a.
%  dataset in
% with the only exception of 
This is because \texttt{census} contains very discriminative attributes, whose
values are employed as blocking keys for \textsf{PSN}\footnote{Soundex
encoded surnames concatenated to initials and zipcodes.},
identifying its duplicates with very high precision.
%\footnote{Please refer to \cite{DBLP:journals/tkde/Christen12} for the definition of the blocking keys for all the datasets.}
Moreover, the profiles of \texttt{census} have short strings as attribute
values:
% , i.e., every profile contains only a few tokens
%, with limited overlap
% most 
on average, every profile contains just 4-5 distinct tokens in its values.
Inevitably, this sparse information has significant impact on the performance of
similarity- and equality-based methods, restricting the co-occurrence
patterns that lie at their core, i.e., the co-occurrences in windows for the
former, and in blocks for the latter.
The impact is larger in the latter case, due to the stricter definition of
co-occurrence, which requires the equality of tokens, not just their similarity.

% the average number of tokens per profile is 5.
% This is the worst case for \textsf{BG}-based methods, which rely on the token overlapping of the duplicates (see \textsf{BG}-methods on \texttt{census} in Figure \ref{fig:pmb_c}a).
% Remember that we are employing \textsf{Token Blocking} to build the starting block collection.
% Similar considerations can be made for \textsf{SN}-based methods in Figure \ref{fig:psn_d}a: they rely on the co-occurrence of the profiles on multiple windows to assign weights to the comparisons, hence, with a limited number of tokens, the co-occurrence in the windows is limited as well.

%%%%%% begin ICDE
%On the other hand, for datasets with a high token overlap of matching profiles,
%and low discriminative attributes, the recall progressiveness of \textsf{PBS}
%is significantly higher than (Figure \ref{fig:small}b-c) or similar to (Figure
%\ref{fig:small}d) that of schema-based \textsf{PSN}.
%%%%%% end ICDE

On the other hand, for datasets with high token overlap between
matching profiles (i.e., they share many attribute value tokens) and non-discriminative 
attributes, which have the same value for many different profiles, our methods
significantly outperform the schema-based \textsf{PSN}.
For instance, the performance of \textsf{PPS} in the {\small \texttt{restaurant}} dataset (Figure
\ref{fig:small}b) is very close to the ideal method: {\small
$AUC_{PPS}^*@1$=0.93}, i.e., 104 out of the first 112 emitted comparisons
are matches\footnote{104 is 93\% of 112, which is the number of existing duplicates in \texttt{restaurant}.}.

%our methods significantly outperform the schema-based \textsf{PSN}.
%For instance, in the \texttt{restaurant} dataset, the performance of
%\textsf{PPS} is very close to the ideal method: $AUC_{PPS}^*@1 = 0.93$, meaning
%that 104 out of the first 112 emitted comparisons are matches\footnote{104 is
%93\% of 112, which is the number of existing duplicates in
%\texttt{restaurant}.}.

%%%%%%%%%%%%%%%%%%%%
%
%   note on the curved of Figure 9
%
%%%%%%%%%%%%%%%%%%%%
%In Figure \ref{fig:small}, our advanced methods shows more constant behaviour among different datasets than PSN, which shows a different progressive behaviour for each datasets.
%In fact: on \texttt{census} the its curve stats with a steep slope, which constantly decreases;
%on \texttt{restaurant}, the slope of its curve increases slowly until $ec^*=8$, steeply from $ec^*=8$ until $ec^*=15$, when sharply decreases;
%on \texttt{cora}, the slope of its curve increases steeply until $ec^*=5$, and then decreases sharply.
%These differences at large extent depend on the schema-based blocking keys definition, which is out of the scope of this paper (recall that we are employing best blocking keys known from \cite{}).
%\note{any conclusion we can state?}
%%%%%%%%%%%%%%%%%%%%
%%%%%%%%%%%%%%%%%%%%
%%%%%%%%%%%%%%%%%%%%

Among the advanced methods, we now list the best performer for each dataset.
On \texttt{census} (Figure \ref{fig:small}a), \textsf{GS-PSN} is the best
performer, but \textsf{LS-PSN} is only slightly worse.
%%%%% begin ICDE
%On \texttt{restaurant} (Figure \ref{fig:small}b) and \texttt{cddb} (Figure \ref{fig:small}d), \textsf{LS-PSN} has the best
%recall progressiveness.
%%%%% end ICDE
On \texttt{restaurant} (Figure \ref{fig:small}b), \textsf{PPS} has the best progressiveness until recall 98\%, but \textsf{LS-PSN} has a similar progressiveness and reaches 100\% earlier than \textsf{PPS} (due to the plot scale, though, this is not evident in Figure \ref{fig:small}).
%until recall 98\%, but \textsf{LS-PSN} has a similar
%progressiveness and reaches 100\% earlier than \textsf{PPS} (due to the
%scale, though, this is not evident in Figure \ref{fig:small}).
On \texttt{cora} (Figure \ref{fig:small}c), \textsf{GS-PSN} has the best initial
progressiveness, but equality-based methods reach the highest recall from $ec^*{=}4$ on|note that the final recall of \textsf{PBS} and \textsf{PPS} is lower than 100\%, because the
underlying Token Blocking cannot identify all duplicates in \texttt{cora}.
On \texttt{cddb} (Figure \ref{fig:small}d), \textsf{PPS} has the best progressiveness for recall up to 65\%, but for higher recall, \textsf{LS-PSN} is the best performer.

%On \texttt{cddb} (Figure \ref{fig:small}d), \textsf{PPS} has the best
%progressiveness for recall up to 65\%, but for higher recall, \textsf{LS-PSN} is
%the best performer.

%To understand which method is best choice for 
%\textsf{ER} over structured datasets, 
We now compare all the methods with respect to their
mean value of normalized area under the curve.
% of $AUC_m^*$ for all methods.
Figure~\ref{fig:auc_small} shows the mean $AUC^*$ of all methods across all
structured datasets for four different values of $ec^*$: 1, 5, 10 and 20.
We observe that, on average, for any level of $AUC^*$, \textsf{LS-PSN} and
\textsf{GS-PSN} are the top performers, in particular for the earliest phase
of \textsf{Progressive ER}: their $AUC^*@1$ is three times the $AUC^*@1$ of
\textsf{PSN} and \textsf{PBS}, and ${\sim}18\%$ higher than that of
\textsf{PPS}.
% $AUC^*_{PPS}@1$}.

Overall, we conclude that the best performing methods for structured datasets
are {\small \textsf{LS-PSN}} and {\small \textsf{GS-PSN}} (the difference in
their performance is insignificant\footnote{Employing the t-test for assessing the significance of the difference of the means: p-value $=0.95$.}.
{%\color{red}
Thus, the selection of one method over the other should be driven by the
differences in their space and time complexities for the initialization and
emission phases, depending on $w_{max}$.
}
%(see Appendix~\ref{sec:complexities} for the complete discussion on complexities)
The higher $w_{max}$ is, the higher gets the space complexity of
{\small \textsf{GS-PSN}} in comparison to {\small \textsf{LS-PSN}};
% , $O($$\cdot$$\kpp$$\cdot$$|P|)$ and $O(w_{max}$$\cdot$$\kpp$$\cdot$$|P|)$
% respectively;
thus, {\small \textsf{LS-PSN}} should be preferred when the availability of
memory may be a issue.
Yet, if memory is not an issue, {\small \textsf{GS-PSN}} should be
preferred, as it avoids multiple emissions of the same comparisons.
%(which correspond to superfluous invocation of match functions).

%{\color{red}??? SHOULD BE IN ACCORD TO SECTION V (GS-PSN MOTIVATION) ???its constant-time emission phase}.
% $O(1)$, while the time complexity of the \textsf{LS-PSN} emission phase is
% $O($$\kpp$$\cdot$$|P|)$ under some circumstances.

%the performance of Blocking Graph-based methods has a higher variance than that
%of Profile List-based methods in Figure~\ref{fig:auc}b, regardless of $ec^*$.
%This is caused by the stricter definition of co-occurrence patterns they
%employ (i.e., blocking key equality).
%Overall, we conclude that if the profiles are represented by a large
%number of tokens (${>}10$ in our experiments), the best performing method is
%\textsf{CEP-pList}. Otherwise, in the presence of a limited number of tokens per profile, the
%best choices are \textsf{GS-PSN} and/or \textsf{LS-PSN} with \textsf{RCF}.

\begin{figure*}[ht!]
	\centering
	\includegraphics[width=0.99\textwidth]{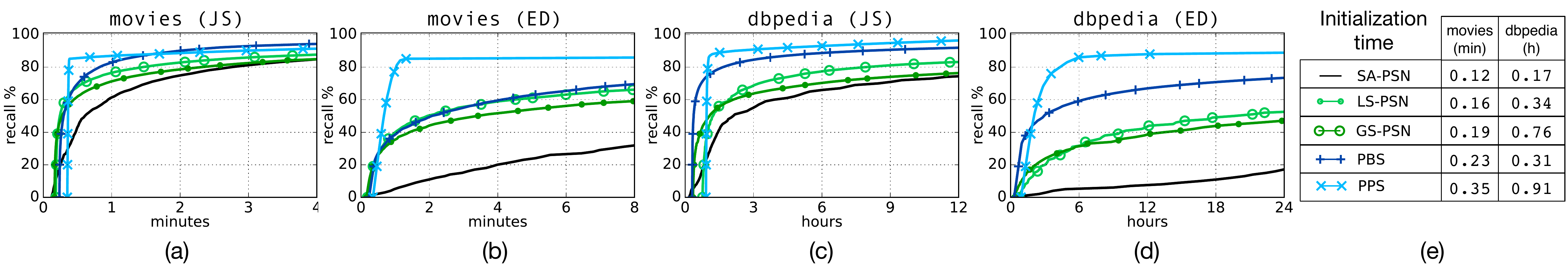}
	\vspace{-8pt}
	\caption{Time experiments with \textsf{jaccard-sim} (a,c) and \textsf{edit-dist} (b,d); initialization times (e).
	}
	\vspace{-14pt}
	\label{fig:time}
\end{figure*}

\vspace{-7pt}
%\subsection{Effectiveness Evaluation on Large and Heterogeneous Datasets}
\subsection{Large, Heterogeneous Datasets}
\label{sec:large}
\vspace{-2pt}

We now assess the relative performance of all methods with respect to recall
progressiveness over the large, heterogeneous datasets \texttt{movies},
\texttt{dbpedia}, \texttt{freebase}.
The corresponding plots appear in Figure~\ref{fig:large}.

The results confirm our intuition about the ineffectiveness of the na\"ive \textsf{SA-PSN} and \textsf{SA-PSAB}, since
all advanced methods outperform it to a significant extent across all datasets.
\textsf{SA-PSAB} also cannot scale to the largest datasets (see Figure \ref{fig:large}b-c) due to the huge blocks in the highest layers of its suffix trees, which entail too many comparisons.

The only exceptions are \textsf{LS-PSN} and \textsf{GS-PSN}\footnote{On
\texttt{freebase}, we limited the number of maximum comparisons of
\textsf{GS-PSN} according to the available memory, i.e., 80GB.} on
\texttt{freebase} (Figure~\ref{fig:large}c), which perform poorly: the performance of \textsf{LS-PSN} 
is similar to that of \textsf{SA-PSN}, while \textsf{GS-PSN} has lower recall
progressiveness than \textsf{SA-PSN}, terminating before achieving a recall
greater that 20\%.
The performance of these two advanced methods can be explained by the
characteristics of the dataset. \texttt{Freebase} is composed of RDF triples.
The extracted tokens consist of RDF keywords, URI, and other RDF properties,
which generate a noisy Neighbor List, since their alphabetical ordering is often
meaningless.
Thus, the \textsf{RCF} weighting scheme cannot approximate correctly the
similarity of the profiles.
% Better performances might be achieved by applying opportune preprocessing step
% to prepare the data (e.g., filtering only relevant tokens), but it is out of
% the scope of this paper investigate that.
On the other hand, \textsf{PBS} is able to get the most of the
semantics in URI tokens, due to the equality requirement,
% which are based on the equality of the tokens and not on their similarity, 
thus being more robust on \texttt{freebase} than the similarity-based methods.

Overall, \textsf{PPS} is the best performer on \texttt{movies} (Figure~\ref{fig:large}a) and \texttt{dbpedia} (Figure~\ref{fig:large}b); while, on \texttt{freebase} (Figure~\ref{fig:large}c), \textsf{PBS} achieves the highest recall progressiveness for $ec^*{<}2$ and $ec^*{>}12$ (\textsf{PPS} is the best performer for $2{<}ec^*{<}12$).
Again, to understand which method is the top performer, we compare them with respect to their mean value of normalized area under the curve.
Figure~\ref{fig:auc_large} shows the mean $AUC_m^*$ of all methods across all datasets for four different values of $ec^*$: 1, 5, 10 and 20.
We observe that \textsf{PPS} is the best performer for any level of $AUC_m^*$,
and conclude that, overall, it is the best performing progressive method over large, heterogeneous datasets.

\vspace{-8pt}
\subsection{Time Efficiency Evaluation.}
\label{sec:time}
\vspace{-2pt}

We note that our methods are general and decoupled from the match function employed to determine whether two profiles are matching or not.
%\setlength{\skip\footins}{16pt}
%These functions are typically expensive, and generally requires labeled data and machine learning algorithms \cite{} \note{citation}.
Yet, to assess their efficiency 
% benefit of our methods 
in terms of
execution time, we evaluate them in combination with two match functions: 
%based on the Damerau-Levenshtein 
edit distance (\textsf{ED}) \cite{DBLP:conf/acsw/Bard07} and 
%the
Jaccard similarity (\textsf{JS}) \cite{DBLP:books/cu/LeskovecRU14}\footnote{In a real-world scenario, each match function would require a threshold parameter to discriminate between matching and non-matching pairs, on the basis of their edit distance (or Jaccard similarity).
Here, we are only interested in measuring the time performance, not the
effectiveness of the match function; hence, we do not employ any threshold, and
the outcome of the match function is assumed to be identical to the known ground truth.}. %between each profile paris emitted, considered as plain strings.
The former is meant to test the performance of our methods with an \textit{expensive} match function, while the latter with a \textit{cheap} one.
The time complexities of \textsf{edit-distance} and \textsf{jaccard-sim} are $O(s{\cdot}t)$ and $O(s{+}t)$, respectively, where $s$ and $t$ are the lengths of the two strings to be compared (i.e., the two profiles compared with the match function).

The schema-based methods are not considered in this evaluation, since they
inherently require an additional overhead time to select the blocking keys (and
to perform the schema-alignment in the case of \textsf{Clean-clean ER}).
There is a plethora of techniques to perform these two
tasks~\cite{DBLP:conf/icde/MelnikGR02,DBLP:journals/tkde/ShvaikoE13,DBLP:conf/aaai/MichelsonK06},
but it is out of the scope of this work to determine which one is the best,
since our proposed methods do not rely on them.

%In Figure~\ref{fig:time}, we report the result of the time experiments on two of the largest datasets (\texttt{movies} and \texttt{dbpedia}), where the \textsf{ER} process is more time and resource consuming.
In Figure~\ref{fig:time}, we report the result of the time experiments on the
datasets \texttt{movies} and \texttt{dbpedia}. %where the \textsf{ER} process
% is more time and resource consuming.
(We do not consider \texttt{freebase} for this test, because
Entity Matching for Linked Data typically 
%composed of RDF triples would 
requires more advanced, iterative algorithms
like SiGMa
\cite{DBLP:conf/kdd/Lacoste-JulienPDKGG13}.\footnote{State-of-the-art
entity matching methods use string similarity as the a-priori similarity of two entities.
Due to high levels of noise and sparsity, they enrich it with contextual information in the form of
matching neighbors, i.e., entities whose URIs are contained in an entity profile
as attribute values. This process is typically \textit{iterative}, constantly
updating the overall similarity of two entities with the evidence gathered from the
latest matches \cite{DBLP:conf/kdd/Lacoste-JulienPDKGG13}.)} In
particular, Figures~\ref{fig:time}a-d plot the performance of all methods, considering both the initialization time and the comparison time.
We did not plot the execution time for \textsf{SA-PSAB} because it is more than an order of magnitude slower than the other methods.
The initialization times are listed in Figure~\ref{fig:time}e{ and
are independent of the match function.} Note that 
% that for all methods and
% datasets the emission time is negligible (w.r.t. the time required by the match functions), and 
we do not report the emission time, as
%it here: 
it is at least two orders of
magnitude smaller than that required by the match functions to compare two
profiles - this applies to all methods and datasets.

%%%%% begin ICDE
%The results in Figure~\ref{fig:time} clearly show that our advanced methods produce most of the matches much earlier than the baseline, with both the expensive and cheap match functions.
%\textsf{LS-PSN} is able to outperform \textsf{SA-PSN} since the early stages of the process, thanks to its fast initialization phase
%(see left part of Figures~\ref{fig:time}a-d).
%\textsf{GS-PSN} and \textsf{PBS} exhibit a similar behavior with minor exceptions: \textsf{SA-PSN} combined with a cheap matching function reaches the 20\% recall mark faster than \textsf{PBS} on \texttt{movies} (Figure \ref{fig:time}a), and faster than \textsf{GS-PSN} on \texttt{dbpedia} (Figure \ref{fig:time}c).

%Overall, \textsf{PBS} achieves higher level of recall much earlier than similarity-based methods on large and heterogeneous datasets.
%The only exceptions are \textsf{LS-PSN} and \textsf{GS-PSN} on \texttt{movies} (Figure \ref{fig:time}a) with the cheap match function, where both similarity-based methods reach recall $=$ 40\% earlier than \textsf{PBS}, thanks to their lower initialization time.
%This is because the overhead of the \textit{Token Blocking Workflow}, which lies at the core of \textsf{PBS}, becomes negligible when using an expensive match function (Figure \ref{fig:time}b) or when applying it to large datasets (Figure \ref{fig:time}c-d).
%%%%% end ICDE

The results in Figure~\ref{fig:time} clearly show that our advanced methods produce most of the matches much earlier than the baseline,
in combination with both the expensive and the cheap match functions.
Similarity- and equality-based methods show different performance characteristics, though.
The difference in the initialization times between \textsf{PBS}/\textsf{LS-PSN} and the baseline is negligible for both match functions (see the left part of Figures~\ref{fig:time}a-d); hence, they are able to outperform the baseline since the early stages of the process.
For \textsf{PPS}, the same consideration is valid only in case the expensive match function is employed (see the left part of Figures~\ref{fig:time}b,d).
In fact, when the cheap match function is employed, its initialization time
(55 minutes over DBPedia) may slightly affect the early stages of
the process (Figures~\ref{fig:time}a,c);

% Overall, the equality-based methods achieve higher level of recall much earlier than similarity-based ones on large and heterogeneous datasets.
% The results show that \textsf{PBS} is suited for \textsf{ER} tasks involving cheap match functions and with very limited time budget ({\color{red}its initialization time is the lowest among the advanced methods}).
% Otherwise, \textsf{PPS} achieves the best performance, both in terms of recall progressiveness (Figure~\ref{fig:auc_large}) and execution time (Figure~\ref{fig:time}).

% \subsection{Considerations}
% {\color{red}
% The experiments demonstrate that \textsf{PPS} exhibits a quite robust performance across both structured and semi-structured (heterogeneous) datasets.
% The same applies to the other equality-based method, \textsf{PBS}, even though it consistently underperforms \textsf{PPS}.
% In contrast, the similarity-based techniques \textsf{LS-PSN} and \textsf{GS-PSN} achieve very high performance over structured datasets and very low one over semi-structured datasets.
% We can conclude, therefore, that equality-based techniques perform well under all settings, while similarity-based techniques can only be used over structured datasets.
% The reason is that these datasets are usually curated, principally containing character-level errors, whereas the semi-structured datasets abound in both character- and token-level noise (e.g., URLs as attribute values).
% In the latter cases, it is harder for two matching entities with similar attribute values to be placed in consecutive positions.}

\vspace{-5pt}
\section{Conclusions and Future Work}
\label{sec:conclusion}
\vspace{-2pt}

We have introduced \textit{schema-agnostic} methods to maximize the recall
progressiveness of Entity Resolution for \textit{pay-as-you-go}
applications, while addressing the Volume and Variety dimensions of Big Data.
%{\color{blue}
They can be distinguished into equality-based (\textsf{PBS} and 
\textsf{PPS}) and similarity-based methods (\textsf{LS-PSN} and \textsf{GS-PSN}).
Our experimental evaluation with several real, structured datasets
demonstrates that the proposed methods significantly outperform the
schema-based state-of-the-art method in the field, \textsf{PSN}, identifying
most of the matches much earlier.

Our experiments also indicate that both equality-based methods 
%\textsf{PBS} and \textsf{PPS} 
exhibit a quite robust performance across both structured and semi-structured
(heterogeneous) datasets. In contrast, both similarity-based techniques 
%\textsf{LS-PSN} and \textsf{GS-PSN} 
achieve very high performance over structured datasets and very low over
semi-structured datasets. The reason is the structured datasets are usually curated,
principally containing character-level errors, whereas the semi-structured
datasets abound in both character- and token-level noise (e.g., URIs as
attribute values). In the latter cases, it is harder for two matching entities
with similar attribute values to be placed in consecutive positions.
We can conclude, therefore, that the similarity-based techniques can only be
used over structured datasets, while the equality-based techniques perform well
under all settings. In fact, \textsf{PBS} is suited for \textsf{ER} tasks
involving cheap match functions and with very limited time budget (its
initialization time is the lowest among the advanced methods).
Otherwise, \textsf{PPS} achieves the best performance, both in terms of recall
progressiveness (Figure~\ref{fig:auc_large}) and execution time
(Figure~\ref{fig:time}).

An interesting direction for extending our work is to 
examine the massive parallelization of our approach based on existing methods
for parallelizing Sorted Neighborhood
\cite{DBLP:journals/cj/MaDY15,DBLP:conf/sac/MestrePN15} and 
Meta-blocking \cite{DBLP:journals/is/Efthymiou0PSP17} in the context of
MapReduce.
\vspace{-10pt}
 
\bibliographystyle{IEEEtran}
% argument is your BibTeX string definitions and bibliography database(s)
\bibliography{references}
\vspace{-25pt}
\begin{IEEEbiographynophoto}
%[{\includegraphics[width=1in,height=1.25in,clip,keepaspectratio]{images/g}}]
{Giovanni Simonini}
is a postdoctoral researcher at the department of Engineering ``Enzo Ferrari'' of the University of Modena and Reggio Emilia, Italy.
He received the PhD degree in Computer Science from the University of Modena and Reggio Emilia.
His doctoral dissertation won the Best Thesis Award from the IEEE Computer Society Italy Section.
His research focuses on data integration and big data management.
% \end{IEEEbiography}
% \vspace{-10pt}
% \begin{IEEEbiography}[{\includegraphics[width=1in,height=1.25in,clip,keepaspectratio]{images/george.jpg}}]

\vspace{2pt}
\noindent
\textbf{George Papadakis} is an internal auditor of information systems and a
research fellow at the University of Athens. He has also worked at the NCSR "Demokritos",
National Technical University of Athens (NTUA), L3S Research Center and "Athena"
Research Center. He holds a PhD in Computer Science from Hanover University
and a Diploma in Computer Engineering from NTUA. His research focuses on web
data mining.
% \end{IEEEbiography}
% \vspace{-10pt}
% \begin{IEEEbiography}[{\includegraphics[width=1in,height=1.25in,clip,keepaspectratio]{images/themis.jpg}}]

\vspace{2pt}
\noindent
\textbf{Themis Palpanas} is Senior Member of the Institut Universitaire de
France (IUF), and Professor of computer science at the Paris Descartes University (France),
where he is the director of diNo, the data management group. He is the author of
nine US patents, three of which have been implemented in world-leading
commercial data management products. He is the recipient of three Best Paper
awards, and the IBM Shared University Research (SUR) Award. He is serving as
Editor in Chief for BDR Journal, Associate Editor for PVLDB 2019 and TKDE
journal, and Editorial Advisory Board member for IS journal.
% \end{IEEEbiography}
% \vspace{-10pt}
% \begin{IEEEbiography}[{\includegraphics[width=1in,height=1.25in,clip,keepaspectratio]{images/s}}]

\vspace{2pt}
\noindent
\textbf{Sonia Bergamaschi} is full professor of Computer Engineering at the Engineering Department "Enzo Ferrari" of the University of Modena and Reggio Emilia, where
she is leading the database research group "DBGroup".
Her research activity has been mainly devoted to knowledge representation and management in the context of very large databases.
She was coordinator and participant of many ICT European projects.
She has published more than two hundred international journal and conference papers.
She has served on the committees of the main international Database and AI conferences.
She is an ACM Distinguished Scientist and an IEEE senior Member.
\end{IEEEbiographynophoto}

\end{document}